\documentclass[11pt]{article}
\usepackage{fullpage,amssymb,amsmath}
\usepackage[dvips]{epsfig,graphicx}

\def\##1{{\bf #1}}
\def\=#1{\underline{\underline{#1}}}

\def\+#1{\underline{\bf #1}}
\def\*#1{\underline{\underline{\bf #1}}}

\def\_#1{\underline{#1}}

\def\r#1{(\ref{#1})}
\def\l#1{\label{#1}}
\def\c#1{\cite{#1}}

\def\le{\left(}
\def\ri{\right)}
\def\les{\left[}
\def\ris{\right]}
\def\lec{\left\{}
\def\ric{\right\}}

\def\.{\mbox{ \tiny{$^\bullet$} }}

\def\lam{\lambda}

\def\det{\mbox{det}}
\def\adj{\mbox{adj}}
\def\Tr{\mbox{tr}}

\begin{document}

\LARGE
\begin{center}
{\bf On the homogenization of orthotropic elastic  composites by the
strong--property--fluctuation theory}

\vspace{10mm} \large

Andrew J. Duncan\footnote{  E--mail:
Andrew.Duncan@ed.ac.uk.},  Tom G. Mackay\footnote{Corresponding author.
E--mail: T.Mackay@ed.ac.uk.}\\
\emph{School of Mathematics and
   Maxwell Institute for Mathematical Sciences\\ University of Edinburgh, Edinburgh EH9
3JZ, UK}

\vspace{3mm}

 Akhlesh  Lakhtakia\footnote{E--mail: akhlesh@psu.edu}\\
 \emph{ NanoMM~---~Nanoengineered Metamaterials Group\\ Department
  of Engineering Science and Mechanics\\
Pennsylvania State University, University Park, PA 16802--6812, USA}

\end{center}

\vspace{4mm}

\normalsize

\begin{abstract}
The strong--property--fluctuation theory (SPFT) provides a general
framework for  estimating  the constitutive parameters of a
homogenized composite material (HCM). We developed the elastodynamic
SPFT for orthotropic HCMs, in order to  undertake numerical studies.
A specific choice of two--point covariance function~---~which
characterizes the distributional statistics  of the generally
ellipsoidal particles that constitute the  component
materials~---~was implemented. Representative numerical examples
revealed that the lowest--order SPFT estimate  of the HCM stiffness
tensor is
 qualitatively similar to the
estimate provided by the Mori--Tanaka mean--field formalism, but the
differences between the two estimates vary as the orthotropic nature
of the HCM is accentuated. The second--order SPFT provides a
correction to the lowest--order estimate of the HCM stiffness tensor
and density. The correction, indicating effective dissipation due to
scattering loss, increases as the HCM becomes less orthotropic but
decreases   as the correlation length becomes smaller.

\end{abstract}


\section{Introduction}

How do we estimate the effective constitutive properties of
composite materials? This question has long been considered in the
context of acoustics, elastodynamics and electromagnetics
\c{Akhlesh_book,Milton_book}.  An upsurge in interest in this topic
has been prompted by the recent proliferation  of
\emph{metamaterials}, both as theoretical concepts and as physical
entities. An operational definition of a metamaterial is as an
artificial composite material which
exhibits properties that are not exhibited by its component materials or at
least not exhibited to the same extent by its component materials
\c{Walser}. Metamaterials are often exemplified by
homogenized composite materials (HCMs). Typically, metamaterials are
associated with constitutive parameter regimes which have not been
accessible conventionally. For example, in relation to
elastodynamics, metamaterials with negative mass density \c{Mei} and
negative stiffness \c{Lakes_01,Fang} have recently been described,
whereas negatively--refracting metamaterials have been the subject
of intense research activity lately in electromagnetics \c{Rama}.

We focus here  on the effective elastodynamic properties of a
composite material. The HCM considered arises in the
long--wavelength regime from component materials which are generally
orthotropic, viscoelastic and randomly distributed as oriented
ellipsoidal particles.
 Our study is based on the
strong--property--fluctuation theory (SPFT) which~---~by allowing
for higher--order characterizations of the distributional statistics
of the component materials~---~provides a multi--scattering approach
to homogenization \c{Ryzhov}. This distinguishes the SPFT from
certain well--known self--consistent approaches to homogenization
\cite{Hill_1963,Hill_1965,Bud,Sabina_Willis}, although we note that
more sophisticated self--consistent theories have been proposed in
recent years \c{Kim,Kanaun,Wang_Qin}. While the general character of
the SPFT approach to homogenization is reminiscent of
multi--scattering theories \c{Twersky,Linton,Avila}, the SPFT
provides an estimate of the HCM's constitutive parameters whereas
multi--scattering approaches generally provide effective wavenumbers
\c{Datta,Varadan,Maurel}. A distinctive feature of
 the SPFT is that it incorporates a
renormalized formulation which can accommodate relatively strong
variations in the constitutive parameters of the component
materials. This is because the perturbative scheme for averaging the
renormalized equations in the SPFT is based  on parameters which
remain small even when there are strong fluctuations in the
constitutive parameters describing the component materials. In
contrast, conventional variational methods of homogenization
\c{HS_1962,Kroner,Willis,Talbot_Willis,Hashin_83} yield bounds which
are widely separated when there are large differences  between the
constitutive parameters of the component materials.

The SPFT has been widely utilized to estimate the electromagnetic
constitutive parameters of HCMs \c{TK81,Genchev,ML95,spft_form,Cui}.
 Acoustic
\c{Zhuck_acoustics} and elastodynamic \cite{spft_zhuck1} versions of
the theory have also been developed. The general framework for the
elastodynamic SPFT, applicable to linear anisotropic HCMs, was
established in 1999 \cite{spft_zhuck1}, but no numerical studies
have been reported hitherto. In the following we apply this theory
to examine numerically the case wherein the component materials are
generally orthotropic materials which are distributed as oriented
ellipsoidal particles. Prior to undertaking our numerical study, we
derive new theoretical results in two areas:
\begin{itemize}
\item[(i)] in the
implementation of a
 two--point covariance function which characterizes the
 distributions of the component materials, and
 \item[(ii)] in the
 simplification of certain integrals in order to make them
 amenable to numerical computation.
 \end{itemize}
The SPFT estimates of the HCM constitutive parameters are
illustrated by means of  numerical examples, and results are
compared to those provided by the Mori--Tanaka mean--field approach
\cite{Mori-Tanaka,Benveniste}.

\section{Theory} \l{theory_section}

\subsection{Preliminaries}

In applying the elastodynamic SPFT formalism, it is expedient to
adopt both matrix and tensor representations \c{Lakhjcm}. The
correspondence between the two representations is described in
Appendix~A.
 Matrixes are denoted by double underlining and bold
font, while vectors are in bold font with no underlining. Tensors
are represented in normal font with  their components indicated by
subscripts (for $n$th--order tensors, with $n \leq 4$) or subscripts
and superscripts (for eighth--order tensors). All tensor indexes
range from $1$ to $3$. The $pq$th component of a matrix $\*A$ is
written as $\les \, \*A \, \ris_{pq}$, while the $p$th component of
a vector $\#b$ is written as $\les \, \#b \, \ris_p$.
 A repeated index
implies summation. Thus, we have the matrix component $\big[ \, \*A
\cdot \*B \,\big]_{pr}=
 \big[\, \*A \, \big]_{pq}\big[ \, \*B \,
\big]_{qr}$,  vector component $\big[ \, \*A \cdot \#b\big]_{p}=
 \big[ \, \*A \, \big]_{pq} \big[\#b\big]_{q}$,
and scalar $\#a \cdot \#b =  \les \#a \ris_p \les \#b \ris_p$. The
adjoint, determinant and trace of a matrix $\*A$ are denoted by
$\adj \le \, \*A\, \ri$, $\det \le \, \*A \, \ri$ and $\Tr \le \,
\*A \, \ri$, respectively.
The prefixes $\mbox{Re}$ and $\mbox{Im}$ are used to signify
real and imaginary parts, respectively, while $i = \sqrt{-1}$.

The  SPFT is developed in the frequency domain wherein the stress,
strain, and displacement have an implicit $\exp \le - i \omega t
\ri$ dependency on time $t$, $\omega$ being the angular
frequency. Thus,
  these   are generally complex--valued quantities.
In order to retrieve the corresponding time--domain quantities, the
inverse temporal Fourier transform operation must be performed,
although one must bear in mind that homogenization is essentially a
long--wavelength procedure \c{Akhlesh_book_intro,M08}.
   The possibility of
viscoelastic behaviour is accommodated through complex--valued
constitutive parameters. Stiffness tensors
are taken to exhibit the usual symmetries
\begin{equation}\label{eq_symmetry}
C_{lmpq} = C_{mlpq} = C_{lmqp} = C_{pqlm},
\end{equation}
whilst noting that the symmetry $\mbox{Im}\, C_{lmpq} = \mbox{Im} \,
C_{pqlm}$
  has not been proved generally \cite{Cerveny_2006}.
On account of the symmetries \r{eq_symmetry},
 the  matrix counterpart of tensor
$C_{lmpq}$~---~namely,
 the 9$\times$9
stiffness matrix $\*C$~---~is symmetric.\footnote{Alternatively, in
light of \r{eq_symmetry}, the stiffness tensor may  be represented
by a symmetric 6$\times$6 matrix \c{Ting}, but the following
presentation of the SPFT is more straightforwardly presented in
terms of the 9$\times$9 matrix representation.}

\subsection{Component materials}

We consider the homogenization of a two--component composite material. The
component materials, which are themselves homogeneous, are  randomly
distributed throughout the mixture as identically--oriented,
conformal, ellipsoidal particles. For convenience, the principal
axes of the ellipsoidal particles are taken to be aligned with the
Cartesian axes. Thus, the surface of each ellipsoidal particle
 may be parameterized by the vector
\begin{equation} \l{r_shape}
\#r^{(e)} = \eta \*U \cdot\hat{\#r},
\end{equation}
where $\eta$ is a linear measure of size, $\hat{\#r}$ is the radial
unit vector and the diagonal shape matrix
\begin{equation}\label{U_shape}
\*U=\frac{1}{\sqrt[3]{abc}}\left(
\begin{array}{ccc}
a & 0 & 0 \\
0 & b & 0 \\
0 & 0 & c \\
\end{array}
\right),\qquad \qquad  (a,b,c \in\mathbb{R}^{+}).
\end{equation}

 Let the space
 occupied by the composite material be
 denoted by $V$. It is partitioned into parts $V^{(1)}$ and
$V^{(2)}$ containing the two component materials labelled as `1' and
`2', respectively. The distributional statistics of the component
materials are described in terms of moments of the characteristic
functions
\begin{equation}
\Phi^{(\ell)} (\#r) = \left\{ \begin{array}{ll} 1, & \qquad \#r \in
V^{(\ell)},\\ & \qquad \qquad \qquad \qquad \qquad \qquad (\ell=1,2) . \\
 0, & \qquad \#r \not\in V^{(\ell)}, \end{array} \right.
\end{equation}
 The volume fraction of component material $\ell$, namely $f^{(\ell)}$ , is given by
the first statistical moment of
 $\Phi^{(\ell)}$ ;
 i.e., \begin{equation} \l{vf}
 \langle \, \Phi^{(\ell)}(\#r) \, \rangle = f^{(\ell)}, \qquad \qquad \le \ell = 1,2  \ri
 ,
 \end{equation}
where the angular brackets denote the ensemble average of the
quantity enclosed.
 Notice that
 $f^{(1)} + f^{(2)} = 1$.
The second statistical moment of $\Phi^{(\ell)}$
 constitutes a two--point covariance function.
The physically--motivated form \c{TKN82}
\begin{equation}
\langle \, \Phi^{(\ell)} (\#r) \, \Phi^{(\ell)} (\#r')\,\rangle =
\left\{
\begin{array}{lll}
\langle \, \Phi^{(\ell)} (\#r) \, \rangle \langle \Phi^{(\ell)}
(\#r')\,\rangle\,, & & \hspace{10mm}  | \, \=U^{-1}\cdot \le   \#r - \#r' \ri | > L \,,\\ && \hspace{25mm} \\
\langle \, \Phi^{(\ell)} (\#r) \, \rangle \,, && \hspace{10mm}
 | \, \=U^{-1} \cdot \le  \#r -
\#r' \ri | \leq L\,,
\end{array}
\right.
 \l{cov}
\end{equation}
is adopted, where $L>0$ is the correlation length which is taken to
be much smaller than the elastodynamic wavelengths but larger than
the sizes of the component particles. In the context of the
electromagnetic SPFT, the specific form of the covariance function
has only a secondary influence on estimates of HCM constitutive
parameters,
 for a range of
physically--plausible covariance functions \c{MLW01b}.

 The elastodynamic properties of the
   component materials `1' and `2' are characterized by
their stiffness tensors $C_{lmpq}^{(1)}$ and $C_{lmpq}^{(2)}$ (or,
equivalently, their 9$\times$9 stiffness matrixes $\*C^{(\ell)}$,
$\ell \in \lec 1,2 \ric$), and their densities $\rho^{(1)}$
and $\rho^{(2)}$. The stiffness tensors exhibit the symmetries
represented in \r{eq_symmetry}. The component materials are  generally
orthotropic \c{Ting} in the following developments;
i.e., the stiffness matrix
for each component material may be expressed as
\begin{equation} \l{ortho_form2}
\*C^{(\ell)} = \le \begin{array}{lll} \*{\mathcal{M}}^{(\ell)} & \*0
& \*0 \\ \*0 & \*{\mathcal{D}}^{(\ell)}  & \*{\mathcal{D}}^{(\ell)}
\\
\*0 & \*{\mathcal{D}}^{(\ell)}  & \*{\mathcal{D}}^{(\ell)}
\end{array} \ri, \qquad \qquad (\ell=1,2),
\end{equation}
where $ \*{\mathcal{M}}^{(\ell)}$ and  $\*{\mathcal{D}}^{(\ell)}$
are symmetric and diagonal 3$\times$3 matrixes, respectively, and
$\*0$ is the 3$\times$3 null matrix. For the degenerate case in
which the component material `$\ell$' is isotropic, we have
\begin{equation} \l{ortho_form}
\left.
\begin{array}{l}
 \les \*C^{(\ell)} \ris_{11} = \les \*C^{(\ell)} \ris_{22} =
\les \*C^{(\ell)} \ris_{33} = \lambda^{(\ell)} + 2 \mu^{(\ell)} \vspace{4pt} \\
 \les \*C^{(\ell)} \ris_{12} = \les \*C^{(\ell)} \ris_{13} =
\les \*C^{(\ell)} \ris_{23} = \lambda^{(\ell)} \vspace{4pt} \\
 \les \*C^{(\ell)} \ris_{44} = \les \*C^{(\ell)} \ris_{55} =
\les \*C^{(\ell)} \ris_{66} =  \mu^{(\ell)}
\end{array}
\right\}, \qquad \qquad (\ell=1,2),
\end{equation}
where $\lambda^{(\ell)}$ and $\mu^{(\ell)}$ are the Lam\'{e}
constants \c{MNLS}.

\subsection{Comparison material}

 A central concept in the SPFT is that of a  homogeneous
 \emph{comparison material}. This  provides the initial ansatz for an iterative
procedure that
 delivers a succession of  SPFT estimates of the
 constitutive properties of the HCM.
As such, the comparison material represents the lowest--order SPFT
estimate of the HCM.
 Since we have taken the component materials
 to be generally orthotropic and
distributed as ellipsoidal particles,  the comparison material is
generally orthotropic\footnote{In fact, the comparison material
would also  be orthotropic if (i) the components materials were
isotropic but distributed as aligned ellipsoidal particles; or (ii)
the components materials were orthotropic but distributed as
spherical particles}. While this is a physically--reasonable
assumption here, we remark that the form of the HCM stiffness tensor
may be derived via certain asymptotic approaches to homogenization
\c{Parnell}.
 The orthotropic comparison material (OCM) is characterized by its
stiffness tensor $C^{(ocm)}_{lmpq}$ and density $\rho^{(ocm)}$, with
$C^{(ocm)}_{lmpq}$ exhibiting the symmetries \r{eq_symmetry}.

The SPFT formulation exploits  the spectral Green function of the
OCM, which may be expressed in 3$\times$3 matrix form as
\begin{equation}\label{G_eq}
\*G^{(ocm)}(\#k) = \Big[k^2
\*a(\#{\hat{k}})-\omega^2\rho^{(ocm)}{\*I} \Big]^{-1},
\end{equation}
with $\*I$ being the 3$\times $3 identity matrix and
$\*a(\#{\hat{k}})$  the $3 \times 3$ matrix with entries
\begin{equation}\label{aform}
\big[\,\*a (\#{\hat{k}}) \,\big]_{mp}=\frac{k_l C_{lmpq}^{(ocm)}
k_q}{k^2}.
\end{equation}
Herein,  $\#k = k \#{\hat{k}}$  $ \equiv \le k_1, k_2, k_3 \ri$ with
$ \#{\hat{k}} = ( \sin \theta \cos \phi$, $\sin \theta \sin \phi$,
$\cos \theta )$.  For use later on in \S\ref{sec_SPFT}, we remark
that $\*G^{(ocm)} (\#k)$ may be conveniently expressed as
\begin{equation}
\*G^{(ocm)} (\#k) = \frac{1}{ \Delta(\#k)} \, \*N(\#k),
\end{equation}
with the 3$\times$3 matrix function
\begin{equation}
\*N(\#k) =
 k^4 \adj \les \, \*a( \#{\hat{k}} ) \, \ris +\omega^2 \rho^{(ocm)}
k^2 \lec \, \*a( \#{\hat{k}} )-\Tr \les \, \*a( \#{\hat{k}} ) \ris
\, \*I \, \ric+ \le \omega^2 \rho^{(ocm)} \ri^2 \*I
\end{equation}
and the scalar function
\begin{equation} \l{Delta_def}
\Delta(\#k) = k^6 \det \les \, \*a(\#{\hat{k}}) \, \ris -  \omega^2
\rho^{(ocm)} \
 k^4 \Tr
\lec \adj \les \, \*a(\#{\hat{k}}) \, \ris \ric + \le \omega^2
\rho^{(ocm)} \ri^2  k^2 \, \Tr \les \, \*a(\#{\hat{k}}) \ris - \le
\omega^2 \rho^{(ocm)} \ri^3.
\end{equation}

A key step in the SPFT~---~one  which facilitates the calculation of
 $C^{(ocm)}_{lmpq}$ and $\rho^{(ocm)}$~---~
is the imposition of the conditions  \cite[eqs.
(2.72),(2.73)]{spft_zhuck1}
\begin{eqnarray}
&&\left\langle \Phi^{(1)} (\#r) \,\xi_{lmpq}^{(1)} + \Phi^{(2)}
(\#r)\, \xi_{lmpq}^{(2)} \, \right\rangle
 =0, \l{s1}\\
&& \left\langle \Phi^{(1)} (\#r) \le \rho^{(1)}-\rho^{(ocm)} \ri +
\Phi^{(2)} (\#r) \le \rho^{(2)}-\rho^{(ocm)} \ri \right\rangle
 =0 ,\l{s2}
\end{eqnarray}
in order to remove certain secular terms.
In \r{s1}, the quantities
\begin{equation}\label{xi_eq}
\xi_{lmpq}^{(\ell)}=\le C_{lmst}^{(\ell)} - C_{lmst}^{(ocm)} \ri
\eta_{stpq}, \qquad \quad (\ell = 1,2),
\end{equation}
where $\eta^{(\ell)}_{stpq}$ is given implicitly via
\begin{eqnarray}
&&e^{(\ell)}_{pq} = \eta^{(\ell)}_{pqst} f^{(\ell)}_{st}, \l{edef}\\
&&f^{(\ell)}_{ij} = e^{(\ell)}_{ij}+S_{ijlm} \le C_{lmpq}^{(\ell)} -
C_{lmpq}^{(ocm)} \ri e^{(\ell)}_{pq}, \l{fdef}
\end{eqnarray}
and the renormalization tensor
\begin{eqnarray}\label{ellipse_Sint2}
S_{rstu} &=& \frac{1}{8\pi} \int_{0}^{2\pi}d\phi
\int_{0}^{\pi}d\theta \sin \theta \times \nonumber \\ &&
\frac{(\*U^{-1}\cdot \#{\hat{k}})_t \lec (\*U^{-1}\cdot
\#{\hat{k}})_s \big[\, \*a^{-1} (\*U^{-1}\cdot \#{\hat{k}}) \,
\big]_{ru}+(\*U^{-1}\cdot \#{\hat{k}})_r \big[ \, \*a^{-1}
(\*U^{-1}\cdot \#{\hat{k}})\, \big]_{su}
\ric }{(\*U^{-1}\cdot
\#{\hat{k}})\cdot(\*U^{-1}\cdot \#{\hat{k}} )}.
\end{eqnarray}

Upon substituting \r{xi_eq}--\r{fdef} into \r{s1}, exploiting
\r{vf}, and  after  some algebraic manipulations, we obtain
\begin{equation}\label{fullxi}
f^{(1)}\les \le \*C^{(1)}-\*C^{(ocm)} \ri^{\dagger}+\*S\,
\ris^{\dagger} = - f^{(2)}\les \le \*C^{(2)}-\*C^{(ocm)}
\ri^{\dagger}+\*S\, \ris^{\dagger},
\end{equation}
wherein the 9$\times$9 matrix equivalents
 of the  tensors $C_{lmpq}^{(ocm)}$ and $S_{rstu}$ (namely, $\*C^{(ocm)}$ and $\*S$) have been
 introduced and $^{\dagger}$ denotes the matrix  operation defined
in Appendix~A. The
  OCM stiffness matrix may be extracted from \r{fullxi} as
\begin{equation}\label{1stnewit}
\*C^{(ocm)}=\*C^{(1)}+f^{(2)} \les\,
\mbox{\boldmath$\*\tau$}+(\*C^{(2)}-\*C^{(ocm)})\cdot \*S \,
\ris^{\dagger} \cdot\le \*C^{(1)}-\*C^{(2)}\ri,
\end{equation}
where ${\mbox{\boldmath$\*\tau$}}$ is the $9\times 9$ matrix
representation of the identity tensor $\tau_{rstu}$, as described in
Appendix~A. This nonlinear relation \r{1stnewit} can be readily
solved for $\*C^{(ocm)}$ by numerical procedures, such as the Jacobi
method \c{Bagnara}.

By combining \r{vf} with \r{s2}, it follows immediately that the OCM
density is the volume average of the densities of the component
materials `1' and `2'; i.e.,
\begin{equation}
\rho^{(ocm)} = f^{(1)}\rho^{(1)}+f^{(2)}\rho^{(2)}.
\end{equation}

\subsection{Second--order SPFT } \label{sec_SPFT}

The expressions  for the second--order\footnote{The first--order
SPFT estimate is identical to the zeroth--order SPFT estimate which
is represented by the comparison material.} estimates of the HCM
stiffness and density tensors, as derived elsewhere \cite[eqs.
(2.77),(2.78)]{spft_zhuck1}, are
\begin{equation}\label{spft_int}
C^{(spft)}_{lmpq} = C^{(ocm)}_{lmpq}-\frac{\omega^2
\rho^{(ocm)}}{2}\int d^3k \;  \frac{k_t}{k^2}\, B^{lmrs}_{tupq}(\#k)
\, \les \, \*G^{(ocm)} (\#k)\, \ris_{vu} \lec k_s \les \, \*a^{-1}
(\#{\hat{k}})\, \ris_{rv}+k_r \les \, \*a^{-1} (\#{\hat{k}})
\,\ris_{sv} \ric
\end{equation}
and
\begin{equation}\label{rho_int}
\rho^{(spft)}_{mp}= \rho^{(ocm)}\delta_{mp}+\omega^2\int d^3k \;
B(\#k) \les \*G^{(ocm)} (\#k) \ris_{mp},
\end{equation}
respectively,
wherein $\delta_{mp}$ is the Kronecker delta  function. The
eighth--order tensor $B^{lmrs}_{tupq}(\#k)$ and scalar $B(\#k)$
represent the spectral covariance functions given as
\begin{equation}\label{cov_fn}
\left.
\begin{array}{l}
\displaystyle{
 B^{lmrs}_{tupq} (\#k) = \frac{ \le
\xi^{(2)}_{lmrs} - \xi^{(1)}_{lmrs} \ri \le \xi^{(2)}_{tupq} -
\xi^{(1)}_{tupq} \ri}{8 \pi^3} \int d^3 R \;
 \, \Gamma (\#R) \, \exp \le -i \#k \cdot  \#R
\ri} \vspace{4pt}\\
\displaystyle{ B(\#k)= \frac{\le \rho^{(2)}-\rho^{(1)}\ri^2 }{8
\pi^3} \int d^3R \; \Gamma (\#R) \, \exp \le -i \#k \cdot  \#R \ri}
\end{array}
\right\},
\end{equation}
with
\begin{equation}
\Gamma (\#r - \#r') = \langle \, \Phi^{(1)}  (\#r) \, \Phi^{(1)}
(\#r')\,\rangle -  \langle \, \Phi^{(1)}  (\#r) \,\rangle \, \langle
\, \Phi^{(1)} (\#r')\,\rangle \equiv \langle \, \Phi^{(2)} (\#r) \,
\Phi^{(2)} (\#r')\,\rangle - \langle \, \Phi^{(2)}  (\#r) \,\rangle
\, \langle \, \Phi^{(2)} (\#r')\,\rangle.
\end{equation}

We now proceed to simplify the expressions for $C^{(spft)}_{lmpq}$
and $\rho^{(spft)}_{mp}$ presented in  \r{spft_int} and \r{rho_int},
in order to  make them numerically tractable. We
begin with the integral on the right sides of \r{cov_fn} which, upon
implementing
 the step
function--shaped covariance function \r{cov},  may be expressed as
\begin{equation}
\int d^3 R \;
 \, \Gamma (\#R) \, \exp \le -i \#k \cdot  \#R
\ri = \int_{|\#R | \leq L} d^3R \; \exp \les -i \le \*U \cdot \#k
\ri \cdot \#R \ris.
\end{equation}
Thus, we find that $B^{lmrs}_{tupq}(\#k)$ and $B (\#k)$ are given by
\begin{equation}\label{covrfn}
\left.
\begin{array}{l}
\displaystyle{
 B^{lmrs}_{tupq} (\#k) =  \frac{f^{(1)}f^{(2)}
 \le \xi^{(2)}_{lmrs} - \xi^{(1)}_{lmrs} \ri \le
\xi^{(2)}_{tupq} - \xi^{(1)}_{tupq} \ri }{2 \le \pi k \sigma \ri^2}
\les \frac{\sin \le k\sigma L \ri}{k\sigma} -L \cos \le k\sigma L
\ri
\ris} \vspace{4pt} \\
 \displaystyle{ B (\#k) =  \frac{f^{(1)}f^{(2)}
\le \rho^{(2)}-\rho^{(1)}\ri^2   }{2 \le \pi k \sigma \ri^2} \les
\frac{\sin \le k\sigma L \ri}{k\sigma} -L \cos \le k\sigma L \ri
\ris}
\end{array}
\right\},
\end{equation}
wherein the scalar function
\begin{equation}
\sigma\equiv \sigma(\theta,\phi)=\sqrt{a^2 \sin^2\theta
\cos^2\phi+b^2 \sin^2\theta \sin^2\phi +c^2\cos^2\theta}.
\end{equation}

Upon substituting \r{covrfn}  into \r{spft_int} and \r{rho_int}, the
integrals therein with respect to $k$ can be evaluated by means of
calculus of residues:  The roots of $\Delta(\#k) = 0$ give rise to
six poles in the complex--$k$ plane, located at  $k = \pm p_1$, $ \pm
p_2$
 and $ \pm p_3$, chosen such that $\mbox{Re} \; p_i  \geq 0
 \;\; (i=1,2,3)$. From \r{Delta_def}, we find that.
\begin{eqnarray}
  p_1^2 &=& P_A-
\frac{1}{3} \le
  \frac{2^{1/3} P_B}{ P_C \, \det \les \, \*a(\#{\hat{k}}) \, \ris }- \frac{P_C}{  2^{1/3}  \, \det \les \, \*a(\#{\hat{k}}) \, \ris
  }\ri,\\
  p_2^2 &=& P_A+ \frac{1}{3} \le \frac{(1+i\sqrt{3}) P_B}{ 2^{2/3} P_C \, \det \les \, \*a(\#{\hat{k}}) \, \ris
  }-\frac{(1-i\sqrt{3})P_C}{ 2^{4/3}\, \det \les \, \*a(\#{\hat{k}}) \, \ris }\ri, \\
  p_3^2 &=& P_A+
\frac{1}{3} \le
  \frac{(1-i\sqrt{3})P_B}{ 2^{2/3} P_C \, \det \les \, \*a(\#{\hat{k}}) \, \ris
 }-\frac{(1+i\sqrt{3})P_C}{2^{4/3} \, \det
  \les \, \*a(\#{\hat{k}}) \, \ris } \ri,
\end{eqnarray}
wherein
\begin{eqnarray}
  P_A &=& \frac{\omega^2\rho^{(ocm)} \Tr \lec \adj \les \, \*a(\#{\hat{k}}) \, \ris \ric }{3 \,
  \det \les \, \*a(\#{\hat{k}}) \, \ris }, \\
  P_B &=& \le \omega^2 \rho^{(ocm)} \ri^2\le
  3 \, \det \les \, \*a(\#{\hat{k}}) \, \ris \, \Tr \les \, \*a(\#{\hat{k}}) \, \ris
  -\Tr\lec
  \adj \les \, \*a(\#{\hat{k}}) \, \ris \ric^2\ri ,\\
  P_C^3 &=& P_D+\sqrt{4 P_B^3+ P_D^2}, \\
  P_D &=&
\le \omega^2 \rho^{(ocm)} \ri^3
  \Big( 2 \, \Tr \lec \adj \les \, \*a(\#{\hat{k}}) \, \ris \ric^3
  \nonumber \\
  & & -9 \, \det \les \, \*a(\#{\hat{k}}) \, \ris \, \Tr
  \lec
  \adj \les \, \*a(\#{\hat{k}}) \, \ris \ric \, \Tr \les \, \*a(\#{\hat{k}}) \, \ris
  +27 \, \det \les \, \*a(\#{\hat{k}}) \, \ris^2\Big).
\end{eqnarray}
Thus, by application of the
 Cauchy residue theorem \c{Kwok},  the SPFT estimates are delivered as
\begin{eqnarray}\label{final_full_int}
C^{(spft)}_{lmpq} &= & C^{(ocm)}_{lmpq}+\frac{\omega^2 \rho^{(ocm)}
f^{(1)}f^{(2)}\le \xi^{(2)}_{lmrs} - \xi^{(1)}_{lmrs} \ri \le
\xi^{(2)}_{tupq} - \xi^{(1)}_{tupq}\ri}{4\pi i }
\times\nonumber\\
& & \int_{\phi=0}^{2 \pi} \int_{\theta=0}^{\pi} d\phi \, d\theta \,
\frac{k_t  \sin\theta \lec k_s \les \, \*a^{-1}(\#{\hat{k}}) \,
\ris_{rv}+ k_r \les \, \*a^{-1}(\#{\hat{k}}) \, \ris_{sv}\ric}{ \le
k \sigma \ri^2 \; \det \les \, \*a(\#{\hat{k}}) \, \ris }  \les \,
\*b(\#{\hat{k}}) \, \ris_{vu} ,
\end{eqnarray}
and
\begin{eqnarray} \l{rfinal_full_int}
\rho^{(spft)}_{mp}& = & \rho^{(ocm)}\delta_{mp}-\frac{\omega^2
 f^{(1)}f^{(2)}\le
 \rho^{(2)}-\rho^{(1)}\ri^2}{2\pi i }\int_{\phi=0}^{2 \pi}
 \int_{\theta=0}^{\pi} d\phi \,  d\theta
\,  \frac{\sin\theta}{\det \les \, \*a(\#{\hat{k}}) \, \ris}\, \les
\, \*b(\#{\hat{k}}) \, \ris_{mp} ,
\end{eqnarray}
where
\begin{eqnarray}\label{Residue}
\*b(\#{\hat{k}})&=&\frac{1}{2i}\bigg[ \frac{e^{iL\sigma p_1}\*N(p_1
\*U \cdot \#{\hat{k}} )}{\sigma
p_1^2(p_1^2-p_2^2)(p_1^2-p_3^2)}\bigg(1-iL\sigma p_1\bigg) -
\frac{e^{iL\sigma p_2}\*N(p_2 \*U \cdot \#{\hat{k}})}{\sigma
p_2^2(p_1^2-p_2^2)(p_2^2-p_3^2)}\bigg(1-iL\sigma p_2\bigg)\nonumber\\
& &+\frac{e^{iL\sigma p_3}\*N(p_3 \*U \cdot \#{\hat{k}})}{\sigma
p_3^2(p_2^2-p_3^2)(p_1^2-p_3^2)}\bigg(1-iL\sigma p_3\bigg)-
\frac{\*N(\#0)}{\sigma p_1^2p_2^2p_3^2}\bigg].
\end{eqnarray}
The integrals in \r{final_full_int} and \r{rfinal_full_int} are
readily evaluated by standard numerical methods \c{num_methods}.

Significantly, the second--order SPFT estimates $C^{(spft)}_{lmpq}$
and $\rho^{(spft)}_{mp}$
 are complex--valued even when the corresponding quantities for the
 component materials, i.e.,
$C^{(\ell)}_{lmpq}$ and $\rho^{(\ell)}$  ($\ell = 1,2$), are
real--valued.  This reflects the fact that the SPFT effectively
takes into account  losses due to scattering. This feature is not
unique to the SPFT: it arises generally in multi--scattering
approaches to homogenization \c{Twersky,Varadan,Biwa}. We note that
for \cite{Cerveny_2006}
\begin{itemize}
\item[(i)] the time--averaged strain energy density to be
positive--valued,
 $\mbox{Re}\, \*{\breve{C}}^{(spft)}$ is required to be positive--definite; and
\item[(ii)] the time--averaged dissipated energy density to be
positive--valued,
 $ - \mbox{Im}\, \*{\breve{C}}^{(spft)}$ is required to be positive--semi--definite,
\end{itemize}
where $\*{\breve{C}}^{(spft)}$ is the 6$\times$6 matrix with
components $\les \, \*{\breve{C}}^{(spft)} \, \ris_{st} = \les \,
\*C^{(spft)} \, \ris_{st}$ $(s, t \in \lec 1,2,\ldots,6 \ric)$ and
$\*C^{(spft)}$ is the 9$\times$9 matrix equivalent to the SPFT
stiffness tensor $C^{(spft)}_{lmpq}$.

It is  notable too that the second--order  SPFT estimates
$C^{(spft)}_{lmpq}$ and $\rho^{(spft)}_{mp}$ are explicitly
dependent on frequency, whereas the corresponding zeroth--order SPFT
estimates exhibit only an implicit dependency on frequency via the
frequency--dependent constitutive parameters of the component
materials. Accordingly, the second--order SPFT estimates may be
viewed as  low frequency corrections to the quasi--static estimates
provided by the the zeroth--order SPFT.

A complex--valued anisotropic density, as delivered
by \r{rfinal_full_int}, is not without precedent \c{Willis85}; see Milton
\c{Milton_NJP2007} for a discussion on this issue.

\section{Numerical results} \l{Num_Results}

Let us now illustrate the theory presented in \S\ref{theory_section}
by means of some representative numerical examples. We consider
homogenizations wherein the component materials are acetal and glass
(or orthotropic perturbations of these in \S\ref{ortho_perturb}).
The corresponding results are qualitatively similar to those we
found from homogenizations involving a wide range of different
component materials, characterized by widely different constitutive
parameters, which are not presented here.
 In order to
provide a baseline for the SPFT estimate of the HCM stiffness
tensor, the corresponding results provided by the Mori--Tanaka
mean--field formalism \cite{Mori-Tanaka,Benveniste} were also
computed. The Mori--Tanaka formalism was chosen as a comparison for
the SPFT because it is well--established and straightforwardly
implemented \c{Lakhjcm,Mura_book}. Comparative studies involving the
Mori--Tanaka and other homogenization formalisms are  reported
elsewhere; see  \c{Ferrari,Hu_Weng,Mercier}, for example. The
Mori--Tanaka estimate of the 9$\times$9 stiffness matrix of the HCM
may be written as \c{Mura_book}
\begin{equation}\label{MTestimate}
\*C^{(MT)}=\les \, f^{(1)}\*C^{(1)}+f^{(2)}\*C^{(2)}\cdot\*A \,
\ris\cdot\les \, f^{(1)}{\mbox{\boldmath$\*\tau$}}+f^{(2)}\*A\,
\ris^{\dagger},
\end{equation}
where
\begin{equation}
\*A=\les \, {\mbox{\boldmath$\*\tau$}}+\*S^{(esh)}\cdot \le
\*C^{(1)}\ri^{\dagger} \cdot \le \*C^{(2)}-\*C^{(1)} \ri \,
\ris^{\dagger},
\end{equation}
and $\*S^{(esh)}$ is the 9$\times$9  Eshelby matrix \c{Eshelby}. The
evaluation of this matrix is described in Appendix~B.

In the remainder of this section, we present the 9$\times$9 stiffness matrix of the
HCM, namely $\*C^{(hcm)}$, as estimated by the lowest--order SPFT
(i.e., $hcm =ocm$), the second--order SPFT (i.e., $hcm =spft$) and
the Mori--Tanaka mean--field formalism (i.e., $hcm =MT$). The matrix
$\*C^{(hcm)}$ generally has the orthotropic form represented in
\r{ortho_form2} with $\ell = hcm$. We also present the
second--order SPFT density tensor $\rho^{(spft)}_{mp}$; numerical
results for the lowest--order SPFT density  $\rho^{(ocm)}$ need not
be presented here as that quantity is simply the volume average of
the densities of the component materials.
 For all second--order SPFT
computations, we selected $\omega=2\pi\times 10^6$ $\mbox{s}^{-1}$.

\subsection{Isotropic component materials distributed as oriented
ellipsoidal particles} \l{iso_ell}

Let us begin by considering the scenario in which the component materials
are both isotropic. The   component material `1' is taken to be acetal
(i.e., $\lam^{(1)} = \lam^{(ace)}$, $\mu^{(1)} = \mu^{(ace)}$ and
$\rho^{(1)} = \rho^{(ace)}$), and component material `2' to be glass
(i.e., $\lam^{(2)} = \lam^{(gla)}$, $\mu^{(2)} = \mu^{(gla)}$ and
$\rho^{(2)} = \rho^{(gla)}$). The
 Lam\'e constants and densities for these two materials are
 as follows \cite{MacMillan,Shackelford}:
 \begin{equation}\left. \l{glass_acetal}
 \begin{array} {lll}
 \lam^{(ace)}=2.68 \; \mbox{GPa}, & \mu^{(ace)}=1.15 \; \mbox{GPa}, &
\rho^{(ace)} = 1.43 \times 10^{3}\; \mbox{kg} \, \mbox{m}^{-3} \vspace{4pt} \\
\lam^{(gla)}=21.73 \; \mbox{GPa}, &  \mu^{(gla)}=29.2 \; \mbox{GPa},
& \rho^{(gla)} = 2.23 \times 10^{3}\; \mbox{kg} \, \mbox{m}^{-3}
\end{array}
\right\}.
\end{equation}
The eccentricities of the ellipsoidal component
particles are specified by the parameters $\lec a, b, c \ric$,
per \r{r_shape} and \r{U_shape}.

\begin{figure}[h!]
\begin{center}
\resizebox{2.4in}{!}{\includegraphics{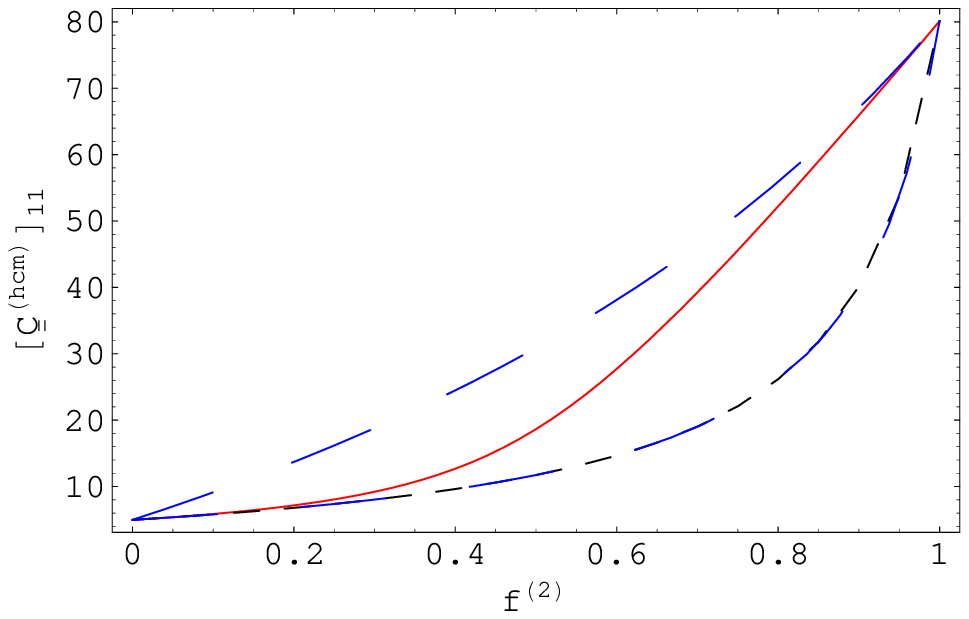}}
\resizebox{2.4in}{!}{\includegraphics{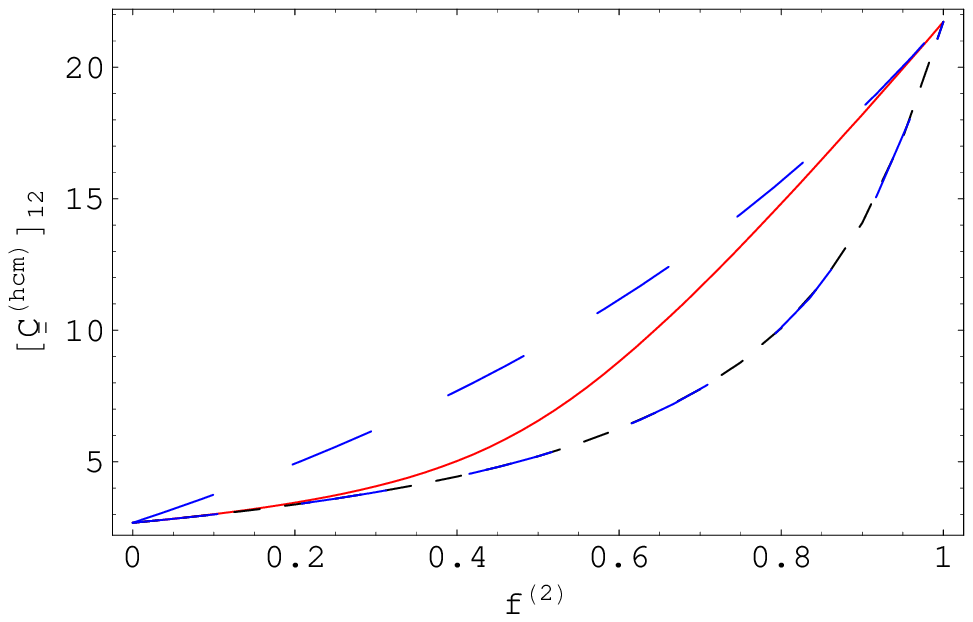}}
\resizebox{2.4in}{!}{\includegraphics{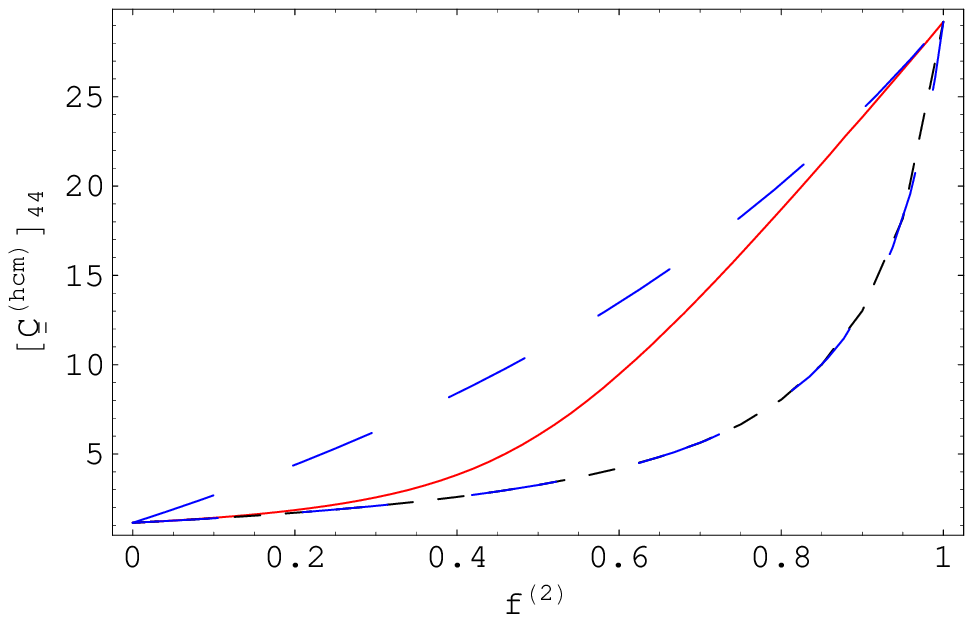}}
\caption{Plots of $\les \, \*C^{(hcm)} \, \ris_{11}$, $\les \,
\*C^{(hcm)} \, \ris_{12}$ and  $\les \, \*C^{(hcm)} \, \ris_{44}$
(in GPa)  as estimated using the lowest--order SPFT (i.e., $hcm =
ocm$) (red, solid curves) and the Mori--Tanaka mean--field formalism
(i.e., $hcm = MT$) (black, dashed curves), against the volume
fraction of component material `2'. Also plotted are the upper and
lower  Hashin--Shtrikman bounds (blue, long dashed curves) for $\les
\, \*C^{(hcm)} \, \ris_{11}$ and $\les \, \*C^{(hcm)} \, \ris_{44}$;
 the lower Hashin--Shtrikman  bounds coincide with the Mori--Tanaka
 estimates.
Component material `1' is acetal and component material `2' is
glass, as  specified in \r{glass_acetal}. The component materials
are distributed as spheres (i.e., $a=b=c$).
}\label{glaaceplot_spher1}
\end{center}
\end{figure}

In  Fig.~\ref{glaaceplot_spher1} the components of the HCM stiffness
matrix $\*C^{(hcm)}$, as computed using the lowest--order SPFT and
the Mori--Tanaka formalism, are plotted as functions of volume
fraction $f^{(2)}$ for the case $a=b=c$. Since the HCM is isotropic
in this case, only the components $\les \, \*C^{(hcm)} \, \ris_{11}
\equiv \lambda^{(hcm)} + 2 \mu^{(hcm)}$, $\les \, \*C^{(hcm)} \,
\ris_{12} \equiv \lambda^{(hcm)}$ and $\les \, \*C^{(hcm)} \,
\ris_{44} \equiv \mu^{(hcm)}$ are presented, per  \r{ortho_form}
with $\ell = hcm$. Notice that the following limits  necessarily
apply  for both the SPFT and Mori--Tanaka estimates:
\begin{equation}
\lim_{f^{(2)} \to 0}  \*C^{(hcm)}
 =  \*C^{(1)},\qquad
\lim_{f^{(2)} \to 1} \*C^{(hcm)}  =
\*C^{(2)}.
\end{equation}
It is apparent from Fig.~\ref{glaaceplot_spher1} that, while the
lowest--order SPFT and the Mori--Tanaka estimates are qualitatively
similar, the Mori--Tanaka estimates display a greater deviation from
the naive HCM estimate $f^{(1)} \les \, \*C^{(1)} \, \ris_{pq} +
f^{(2)} \les \, \*C^{(2)} \, \ris_{pq}$ for mid--range values of
$f^{(2)}$. For further comparison in this isotropic scenario, the
familiar variational bounds on $ \les \, \*C^{(hcm)} \, \ris_{11}$
and $\les \, \*C^{(hcm)} \, \ris_{44}$ established by Hashin and
Shtrikman \c{Milton_book,HS_1962} are also presented in
Fig.~\ref{glaaceplot_spher1}: the lower Hashin--Shtrikman bound
coincides with the Mori--Tanaka estimate and the lowest--order SPFT
estimate lies within the upper and lower Hashin--Shtrikman bounds
for all values of $f^{(2)}$. Parenthetically, we note that for
isotropic HCMs the lowest--order SPFT estimates are the same as
those provided by the well--known formalisms of Hill \c{Hill_1963}
and Budiansky \c{Bud}, as demonstrated elsewhere \c{spft_zhuck1}.

\begin{figure}[h!]
\begin{center}
\resizebox{2.4in}{!}{\includegraphics{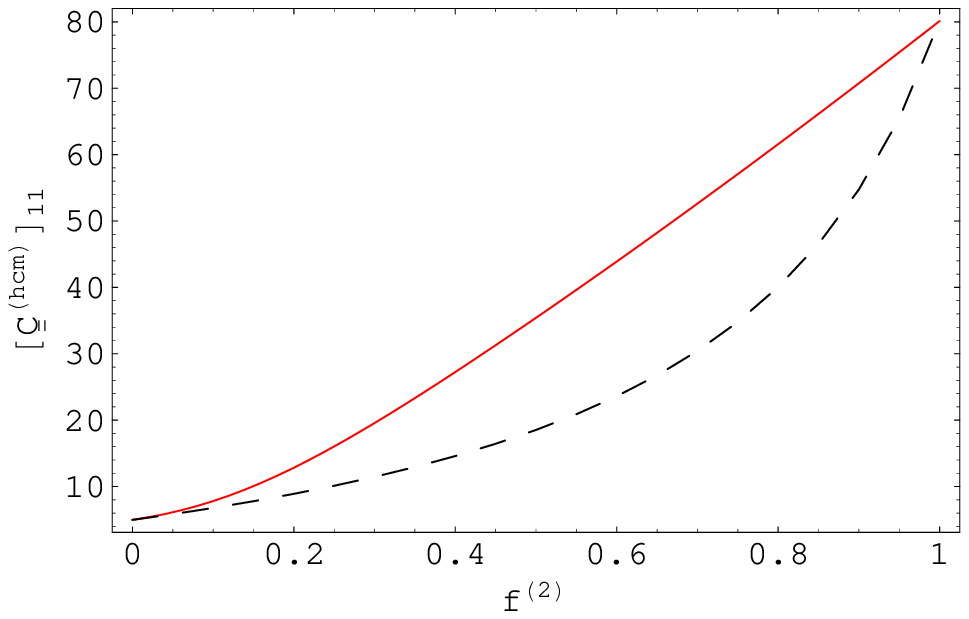}}
\resizebox{2.4in}{!}{\includegraphics{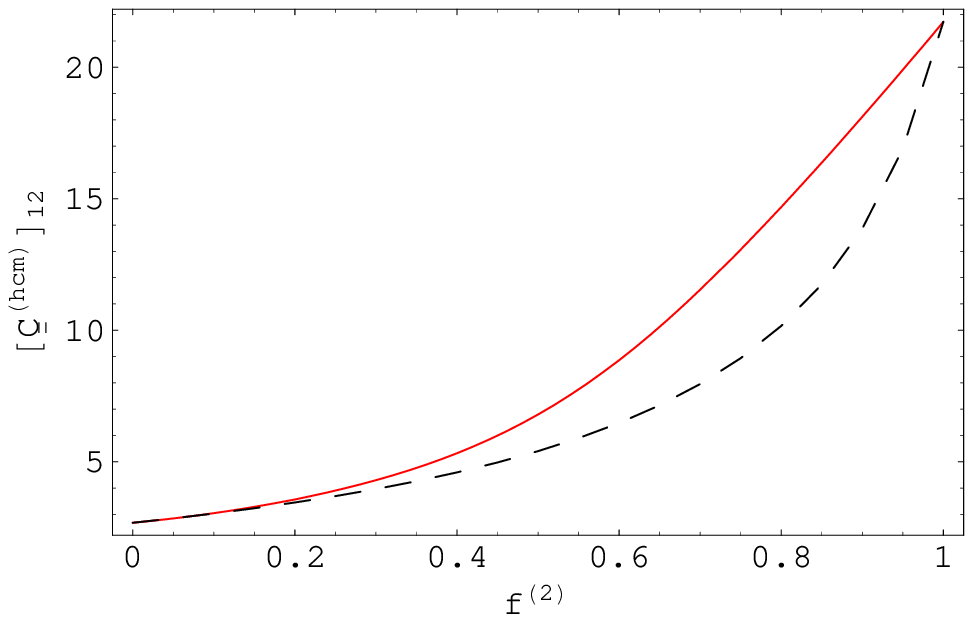}}
\resizebox{2.4in}{!}{\includegraphics{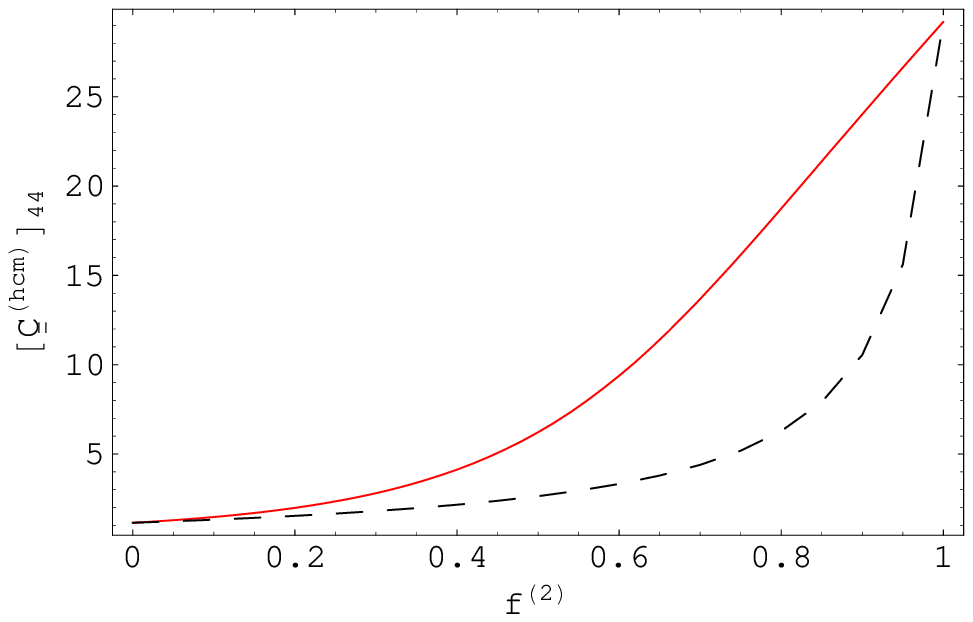}}
\caption{Plots of $\les \, \*C^{(hcm)} \, \ris_{rs}$, with $ rs \in
\lec 11, 12, 44 \ric$ (in GPa) as estimated using the lowest--order
SPFT (i.e., $hcm = ocm$) (red, solid curves) and the Mori--Tanaka
mean--field formalism (i.e., $hcm = MT$) (black, dashed curves),
against the volume fraction of component medium `2'. Component
material `1' is acetal and component material `2' is glass, as
specified in \r{glass_acetal}. The component materials are
distributed as ellipsoids with $a/c = 5$ and $b/c = 1.5$.
}\label{glaaceplot_ellips1}
\end{center}
\end{figure}

The corresponding lowest--order SPFT and Mori--Tanaka estimates  for
the orthotropic HCM arising from the distribution of component
material as ellipsoids  described by $\lec a/c = 5, \, b/c = 1.5
\ric$ are presented in Fig.~\ref{glaaceplot_ellips1}. The matrix
entries $\les \, \*C^{(hcm)} \, \ris_{pq}$ are plotted against
$f^{(2)}$ for $pq \in \lec 11, 12, 44 \ric$. The graphs for $pq =
13$ and $23$ are qualitatively similar to those for $pq = 12$; and
those for $pq = 55$ and $66$ are qualitatively similar to those for
$pq = 44$.
 The degree of orthotropy
exhibited by the HCM can be gauged by  relative differences in the
values of $\les \, \*C^{(hcm)} \, \ris_{pq}$ for $pq \in \lec 11, \,
22,\, 33 \ric$ (and similarly by relative differences in $\les \,
\*C^{(hcm)} \, \ris_{pq}$ for $pq \in \lec 44, \, 55,\, 66 \ric$ and
by relative differences in $\les \, \*C^{(hcm)} \, \ris_{pq}$ for
$pq \in \lec 12, \, 13,\, 23 \ric$). These relative differences are
greatest for mid--range values of the volume fraction $f^{(2)}$.

\begin{figure}[h!]
\begin{center}
\resizebox{2.4in}{!}{\includegraphics{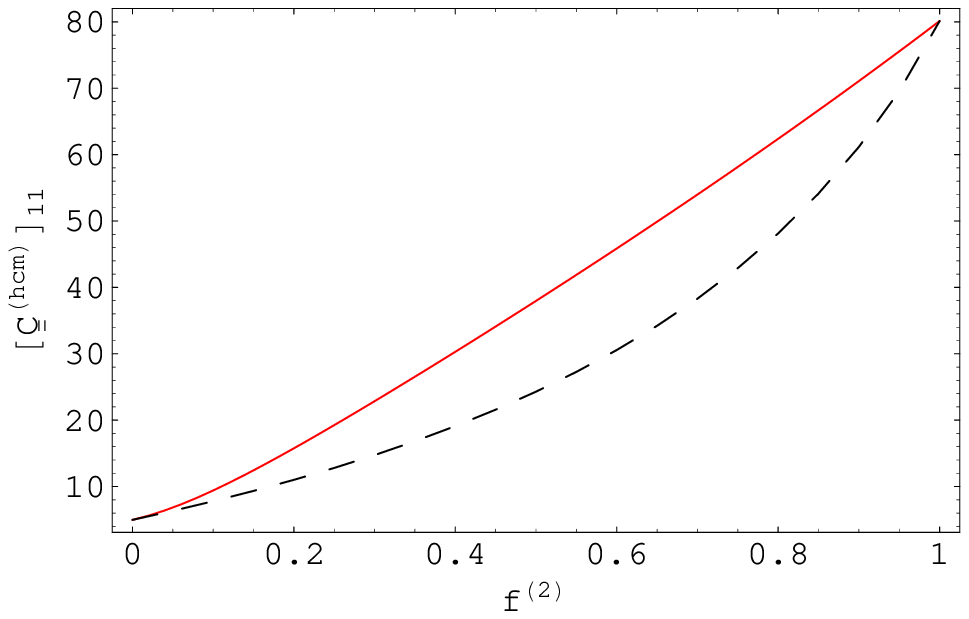}}
\resizebox{2.4in}{!}{\includegraphics{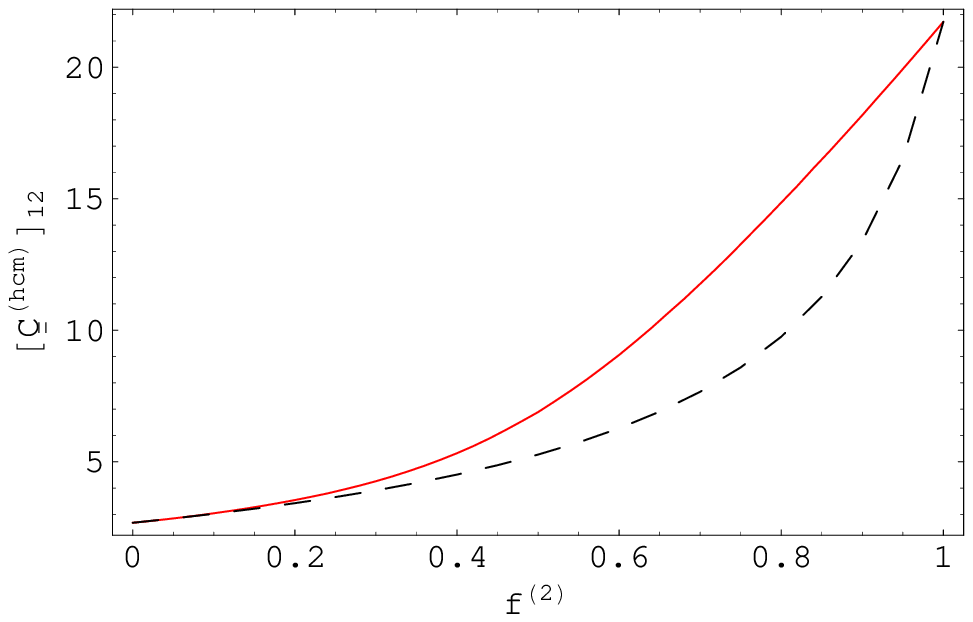}}
\resizebox{2.4in}{!}{\includegraphics{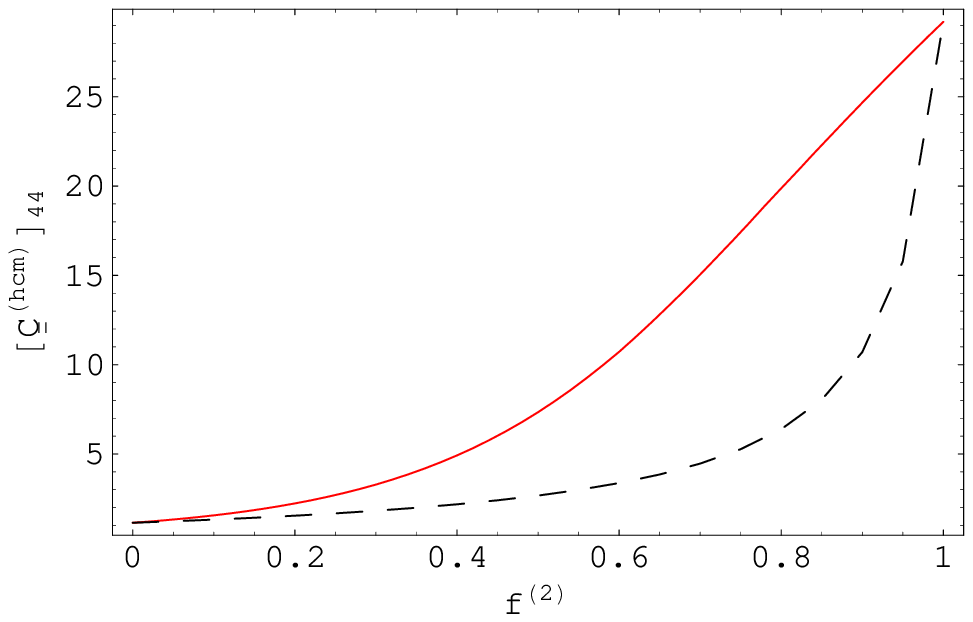}}
\caption{As  Fig.~\ref{glaaceplot_ellips1} but for ellipsoidal
component particles specified by $a/c=10$ and
$b/c=2$.}\label{glaaceplot_ellips2}
\end{center}
\end{figure}

The  orthotropic nature of  the HCM is accentuated by using
component materials with more eccentrically--shaped  particles. This
is illustrated by Fig.~\ref{glaaceplot_ellips2}, which shows results
computed for the same scenario as for Fig.~\ref{glaaceplot_ellips1}
but with ellipsoidal  particles described by $\lec a/c = 10, \, b/c
= 2 \ric$. A comparison of
Figs.~\ref{glaaceplot_spher1}--\ref{glaaceplot_ellips2} reveals that
differences between the estimates of the lowest--order SPFT and the
Mori--Tanaka mean--field formalism  vary slightly as the orthotropic
nature of the HCM is accentuated. For example, the difference
between the  lowest--order SPFT and the Mori--Tanaka estimates of
the $\les \, \*C^{(hcm)} \, \ris_{44}$ increases as the HCM becomes
more orthotropic.

\begin{figure}[h!]
\begin{center}
\resizebox{2.4in}{!}{\includegraphics{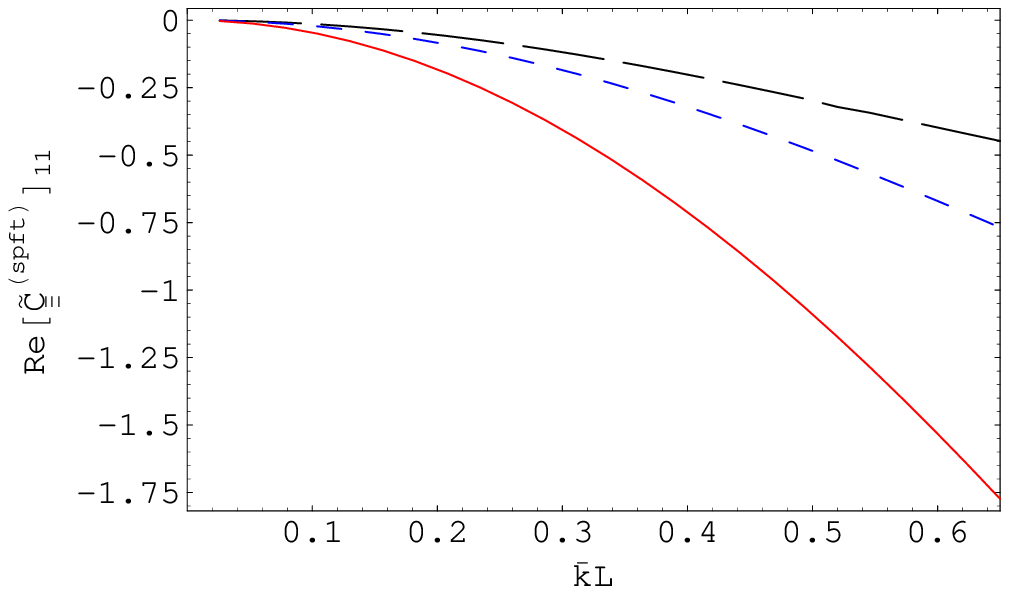}}
\resizebox{2.4in}{!}{\includegraphics{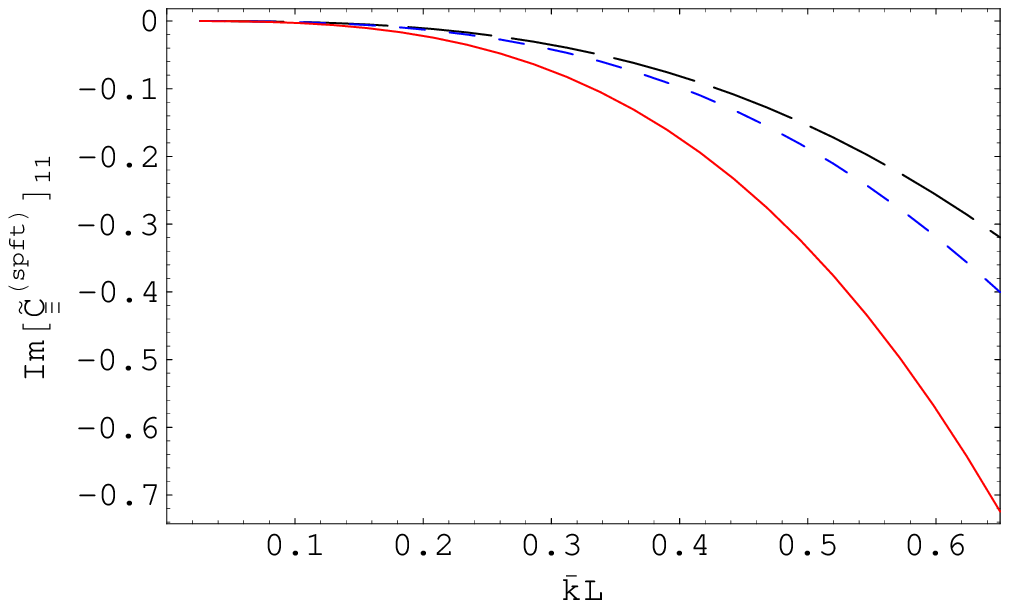}}
\resizebox{2.4in}{!}{\includegraphics{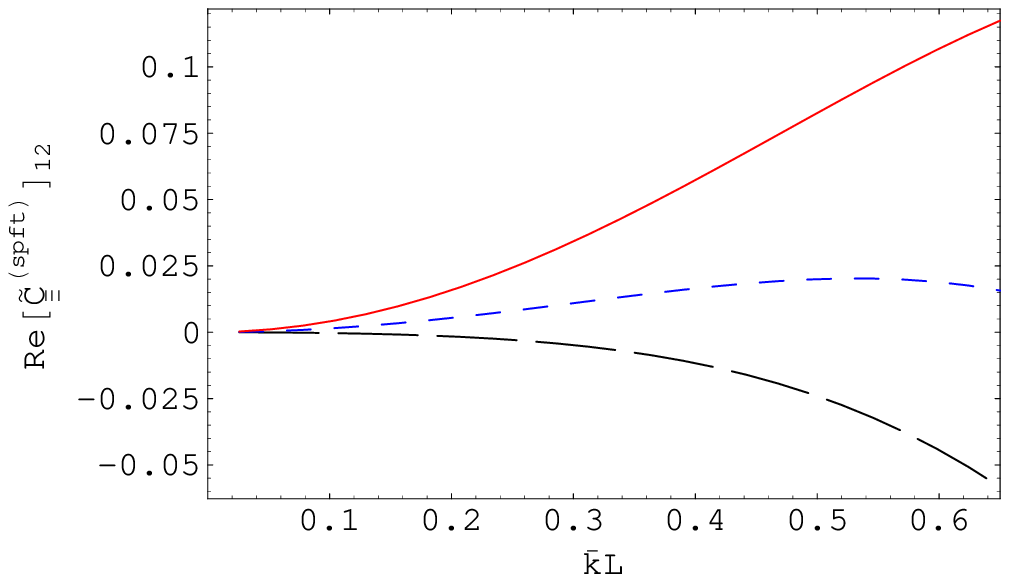}}
\resizebox{2.4in}{!}{\includegraphics{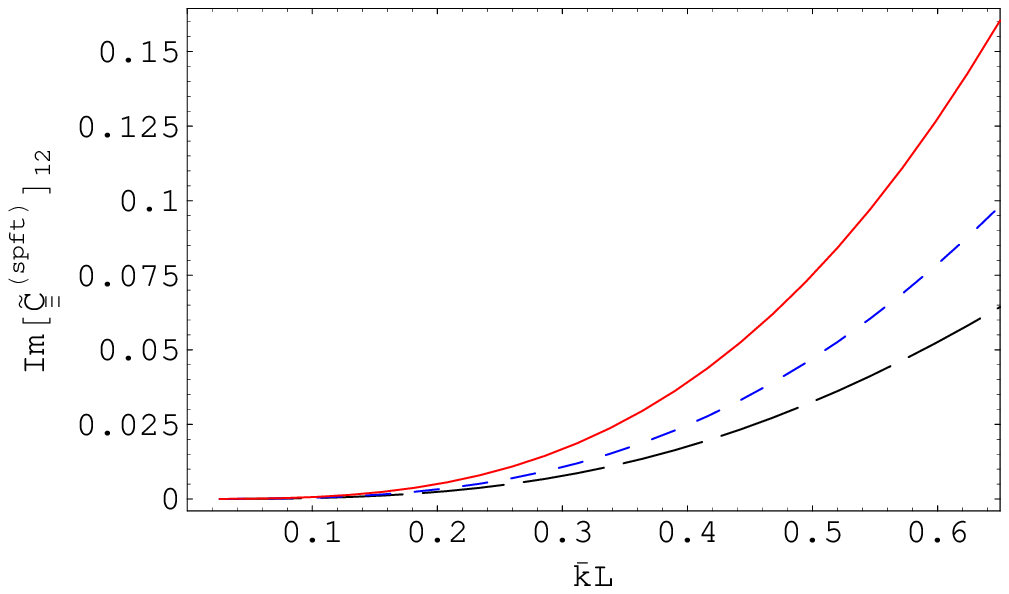}}
\resizebox{2.4in}{!}{\includegraphics{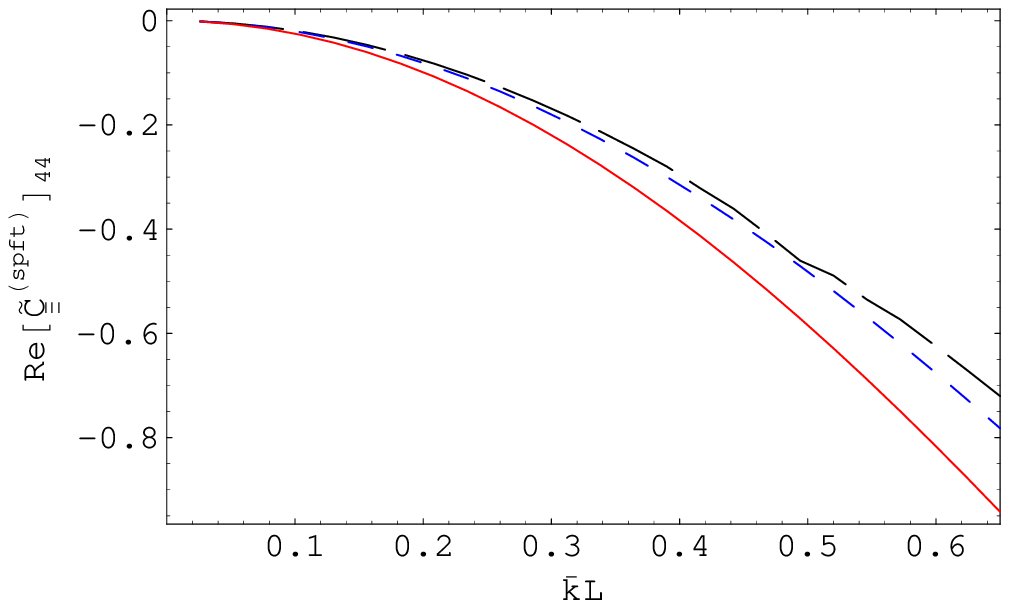}}
\resizebox{2.4in}{!}{\includegraphics{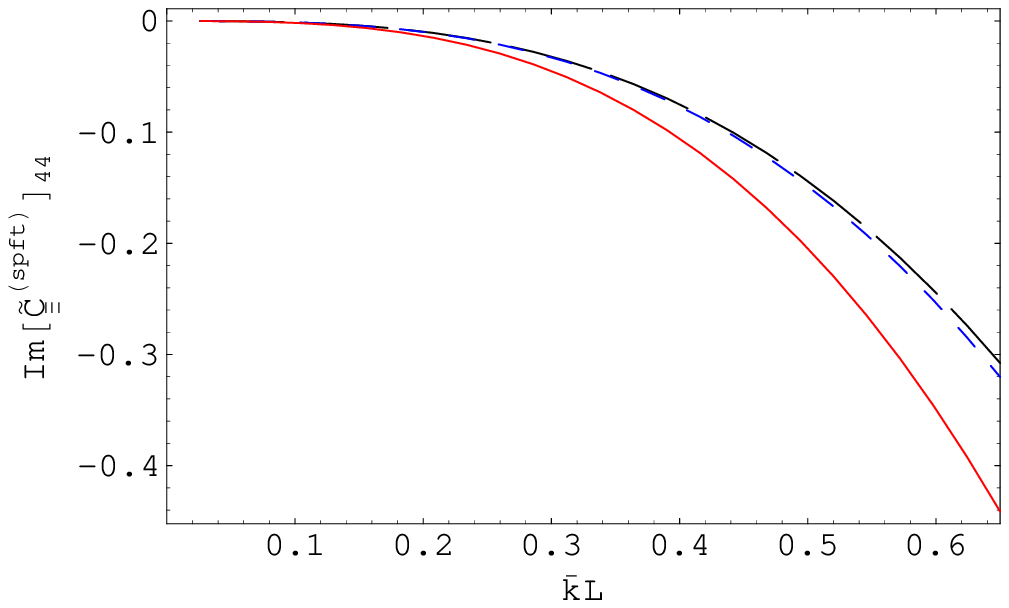}}
\caption{The real and imaginary parts  of $\les
\*{\tilde{C}}^{(spft)} \, \ris_{rs}$, with $ rs \in \lec 11, 12, 44
\ric$ (in GPa), plotted as functions of $\bar{k}L$, for
$f^{(2)}=0.5$. Component medium `1' is acetal and component medium
`2' is glass, as specified in \r{glass_acetal}. The component
materials are distributed as (i) spheres (i.e., $a=b=c$) (red, solid
curves), or (ii) ellipsoids with shape  parameters $\lec a/c = 5, \,
b/c = 1.5 \ric$ (blue, short--dashed curves), or (iii) ellipsoids
with shape parameters $\lec a/c=10, \, b/c=2 \ric$ (black,
long--dashed curves).} \label{Re_Cspftplot_ellipse}
\end{center}
\end{figure}

Now let us turn to the second--order SPFT estimates of the HCM
constitutive parameters. We considered these quantities as functions
of $\bar{k} L$, where
\begin{equation}
\bar{k} = \frac{\omega}{4} \le
\sqrt{\frac{\rho^{(1)}}{\lam^{(1)}+2\mu^{(1)}}} +
\sqrt{\frac{\rho^{(1)}}{\mu^{(1)}}} +
\sqrt{\frac{\rho^{(2)}}{\lam^{(2)}+2\mu^{(2)}}} +
\sqrt{\frac{\rho^{(2)}}{\mu^{(2)}}} \ri
\end{equation}
is an approximate wavenumber  calculated as the average of the shear
and longitudinal wavenumbers in the component materials, and $L$ is
the correlation length associated with the the two--point covariance
function \r{cov}. Since $L$ is required to be smaller than
characteristic wavelengths in the HCM (but larger than the sizes of
the component particles), we restrict our attention to $0 < \bar{k}
L < 0.6$. Fig.~\ref{Re_Cspftplot_ellipse}  shows the real and
imaginary parts of the components of
$\*{\tilde{C}}^{(spft)}=\*C^{(spft)}-\*C^{(ocm)}$ plotted against
$\bar{k} L$ for $f^{(2)} = 0.5$. The values of the shape parameters
$\lec a, \, b, \, c \ric$ correspond to those used in the
calculations for
Figs.~\ref{glaaceplot_spher1}--\ref{glaaceplot_ellips2}. As
previously, only the matrix entries $\les \, \*{\tilde{C}}^{(spft)}
\, \ris_{pq}$ are presented for $pq \in \lec 11, 12, 44 \ric$. The
graphs for $pq = 13$ and $23$ are qualitatively similar to those for
$pq = 12$; and those for $pq = 55$ and $66$ are qualitatively
similar to those for $pq = 44$. Notice that
\begin{equation} \lim_{L \to 0}  \*C^{(spft)} =
 \*C^{(ocm)}
 \end{equation}
and
\begin{equation} \Big\vert \les
\, \*{\tilde{C}}^{(spft)} \, \ris_{pq} \Big\vert
\ll \Big\vert \les \, \*C^{(ocm)} \, \ris_{pq} \Big\vert
\end{equation}
 for all nonzero matrix entries. Furthermore, for the particular example considered here, the
magnitude of $\les \, \*{\tilde{C}}^{(spft)} \, \ris_{pq}$ generally
decreases as the particles of the component materials become more
eccentric in shape.

A very striking feature of the second--order SPFT estimates
presented in  Fig.~\ref{Re_Cspftplot_ellipse} is that
\begin{equation}
\mbox{Im }
  \les \, \*C^{(spft)} \, \ris_{pq} \neq 0,
\end{equation}
 whereas $\mbox{Im } \,
\les \, \*C^{(a,b)} \, \ris_{pq} = \mbox{Im } \, \les \, \*C^{(ocm)}
\, \ris_{pq} = 0$. Furthermore, the magnitude of   $\mbox{Im }\les
\, \*C^{(spft)} \, \ris_{pq}$ grows steadily as the correlation
length increases from zero. These observations may be interpreted in
terms of effective losses due to scattering as follows. For all
reported calculations, $\mbox{Re}\, \*{\breve{C}}^{(spft)}$ is
positive--definite and
 $ - \mbox{Im}\, \*{\breve{C}}^{(spft)}$ is  positive--semi--definite,
which together imply that the associated time--averaged strain
energy and dissipated energy densities are positive--valued
\cite{Cerveny_2006}, as discussed in \S\ref{sec_SPFT}. Accordingly,
the emergence of nonzero imaginary parts of $\les \, \*C^{(spft)} \,
\ris_{pq}$ indicates that the HCM has acquired an effectively
dissipative nature, despite the component materials being
nondissipative. This effective dissipation must be a scattering
loss, because the second--order SPFT accommodates interactions
between spatially--distinct scattering centres via the two--point
covariance function \r{cov}. As the correlation length increases,
the number of scattering centres that can mutually interact also
increases, thereby increasing the scattering loss per unit volume.

\begin{figure}[h!]
\begin{center}
\resizebox{2.4in}{!}{\includegraphics{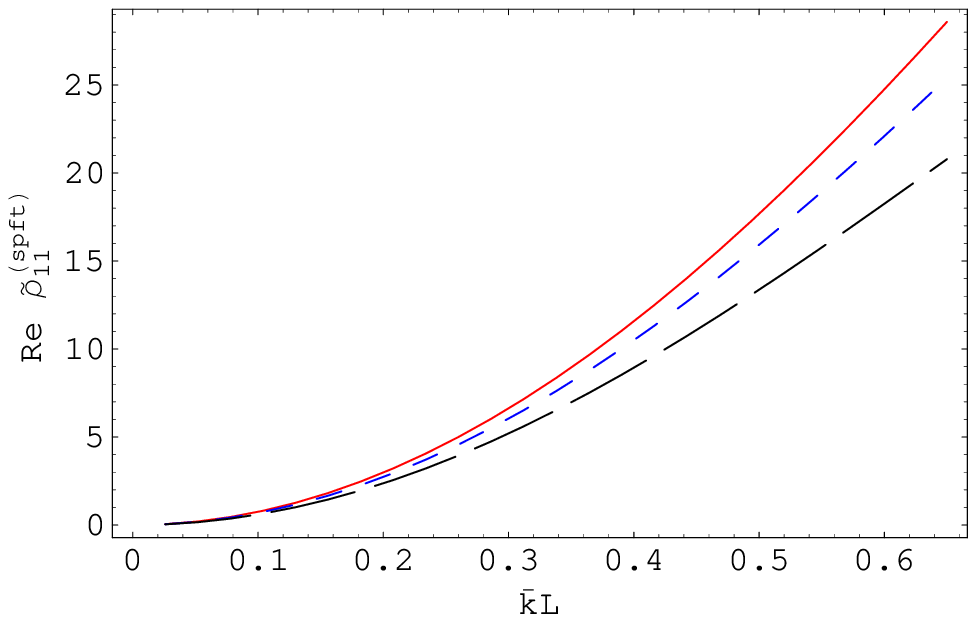}}
\resizebox{2.4in}{!}{\includegraphics{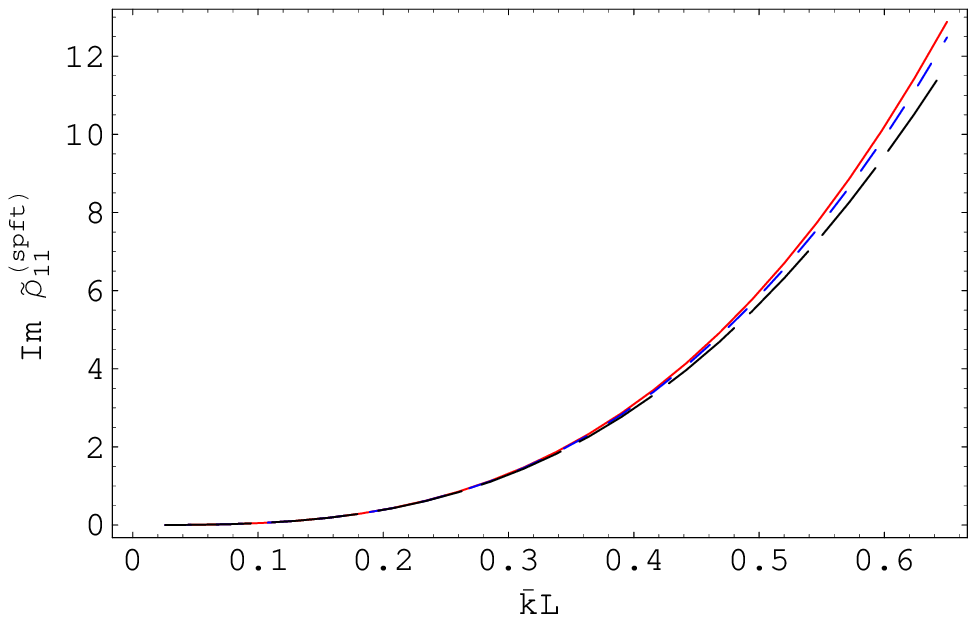}}
\resizebox{2.4in}{!}{\includegraphics{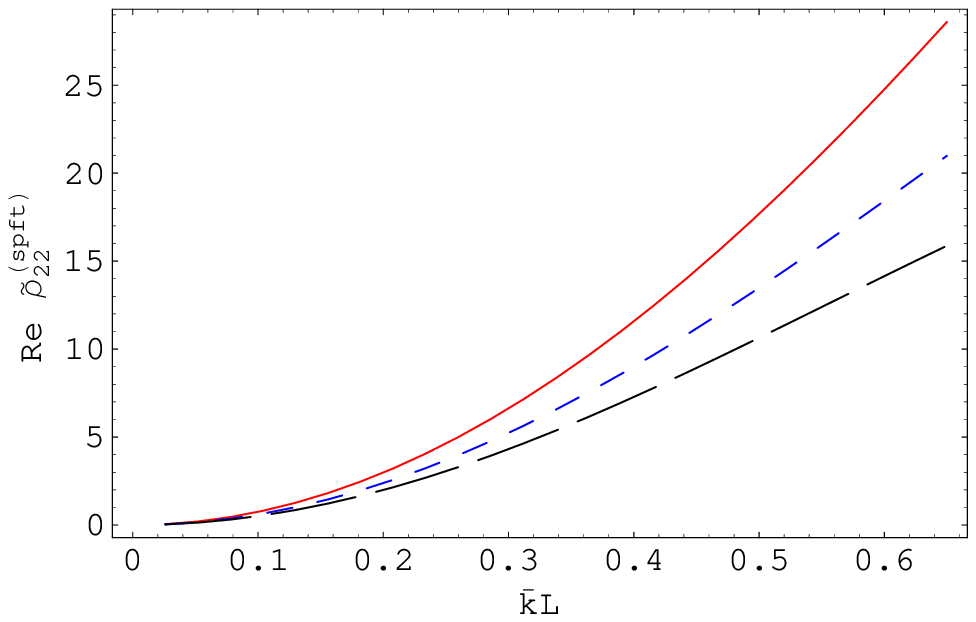}}
\resizebox{2.4in}{!}{\includegraphics{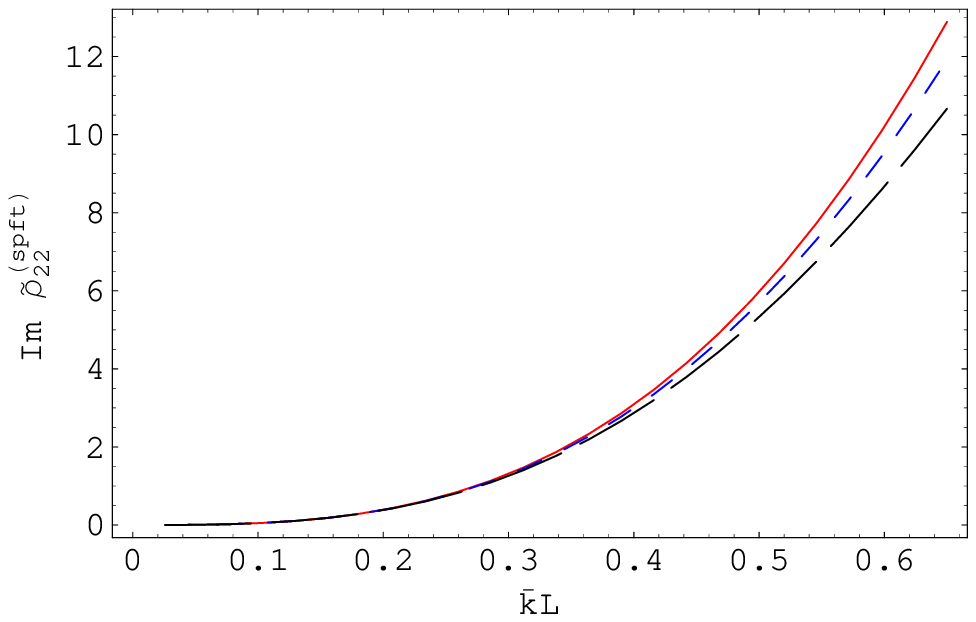}}
\resizebox{2.4in}{!}{\includegraphics{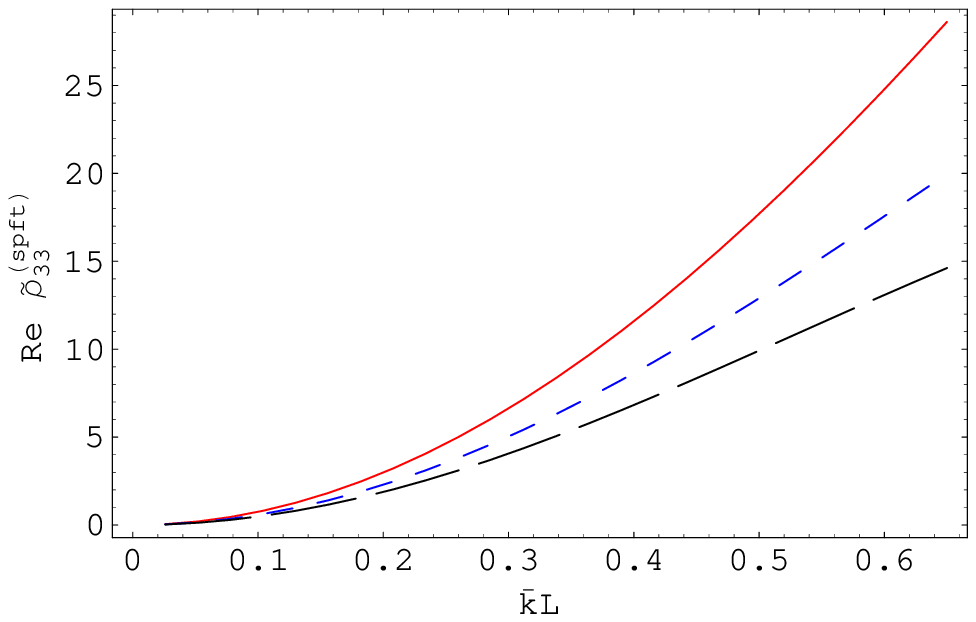}}
\resizebox{2.4in}{!}{\includegraphics{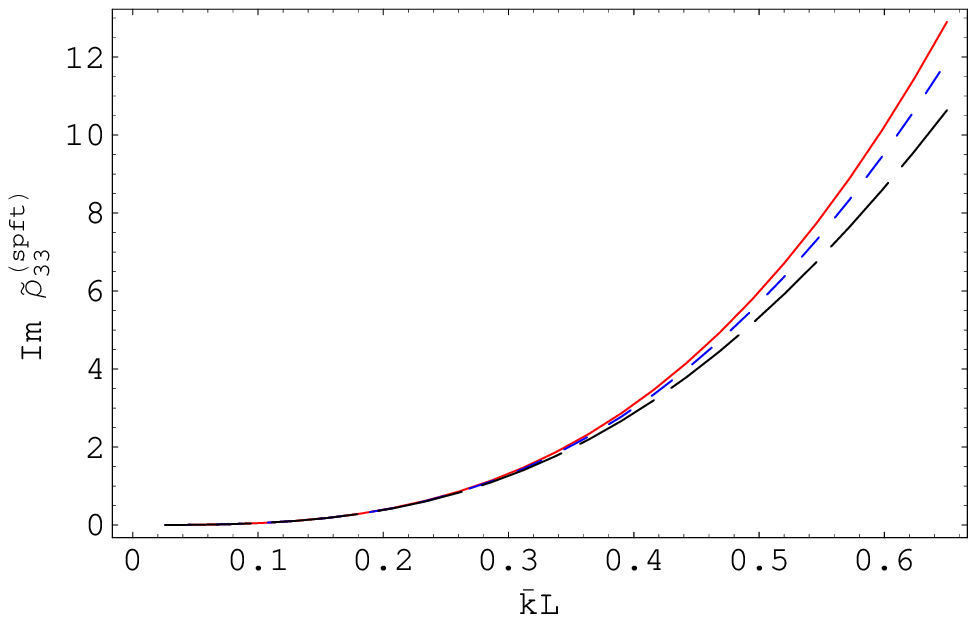}}
\caption{ As Fig.~\ref{Re_Cspftplot_ellipse} but the quantities
plotted are the real and imaginary  parts  of the excess of the
second--order SPFT density tensor over the density of the comparison
material, i.e., $\tilde{\rho}^{(spft)}_{rr} = \rho^{(spft)}_{rr} -
\rho^{(ocm)}$, ($r \in \lec 1,2,3 \ric$), in $ \mbox{kg} \,
\mbox{m}^{-3}$. }\label{Re_rho_ellipse}
\end{center}
\end{figure}

Lastly in this subsection,  the  real and imaginary  parts  of the
second--order SPFT density tensor $\tilde{\rho}^{(spft)}_{pq} =
\rho^{(spft)}_{pq} - \rho^{(ocm)}$ are plotted as functions of
$\bar{k} L$ in  Fig.~\ref{Re_rho_ellipse}. Only the $p=q$ components
are presented, as the $p \neq q$ components  are negligibly small.
The density tensor exhibits characteristics similar to those of the
corresponding stiffness tensor insofar as
\begin{equation}
 \lim_{L
\to 0} \rho^{(spft)}_{pq} =  \rho^{(ocm)}
\end{equation}
and
\begin{equation}
\Big\vert  \tilde{\rho}^{(spft)}_{pq} \Big\vert
 \ll \Big\vert \rho^{(ocm)} \Big\vert
\end{equation}
for all values of
the indexes $p$ and $q$. Also, $ \displaystyle{ \vert
\tilde{\rho}^{(spft)} _{pq} \vert} $ generally decreases as
the shape of the particles of the component materials deviates further from
spherical.

\subsection{Orthotropic component materials distributed as spheres}
\l{ortho_perturb}

Let us now explore the scenario wherein the component materials are
orthotropic perturbations of the isotropic
component materials considered in \S\ref{iso_ell}. In the notation of
\r{ortho_form2}, we choose
\begin{equation}
\left.
\begin{array}{l} \l{om1}
 \*{\mathcal{M}}^{(1)} =
 \le
\begin{array}{lll}
\le \lam^{(ace)} + 2  \mu^{(ace)}\ri \le 1+ \varsigma \ri &
\lam^{(ace)} \le 1-\varsigma \ri & \lam^{(ace)} \le 1 + 2 \varsigma
\ri\\
\lam^{(ace)} \le 1-\varsigma \ri & \le \lam^{(ace)} + 2
\mu^{(ace)}\ri  \le 1 - \frac{1}{4}\varsigma \ri & \lam^{(ace)} \le
1+ \frac{1}{4} \varsigma \ri
\\
\lam^{(ace)} \le 1+2\varsigma \ri & \lam^{(ace)} \le 1 +
\frac{1}{4}\varsigma \ri & \le \lam^{(ace)} + 2  \mu^{(ace)}
\ri
 \le
1- 2 \varsigma \ri
\end{array}
\ri \vspace{8pt}
\\
\*{\mathcal{D}}^{(1)} =
 \le
\begin{array}{lll}
\le   \mu^{(ace)}\ri \le 1- \varsigma \ri & 0 &
0 \\
0 &  \mu^{(ace)} \le 1 - \frac{1}{2}\varsigma \ri &  0
\\
0 & 0 &  \mu^{(ace)}
 \le
1- \frac{2}{3} \varsigma \ri
\end{array}
\ri
\end{array}
\right\}
\end{equation}
and
\begin{equation}
\left.
\begin{array}{l} \l{om2}
 \*{\mathcal{M}}^{(2)} =
 \le
\begin{array}{lll}
\le \lam^{(gla)} + 2  \mu^{(gla)}\ri \le 1+ 2\varsigma \ri &
\lam^{(gla)} \le 1-2\varsigma \ri & \lam^{(gla)} \le 1 + \frac{1}{2}
\varsigma
\ri\\
\lam^{(gla)} \le 1-2 \varsigma \ri & \le \lam^{(gla)} + 2
\mu^{(gla)}\ri  \le 1 + \frac{1}{3}\varsigma \ri & \lam^{(gla)} \le
1- \frac{1}{3} \varsigma \ri
\\
\lam^{(gla)} \le 1+ \frac{1}{2}\varsigma \ri & \lam^{(gla)} \le 1 -
\frac{1}{3}\varsigma \ri & \le \lam^{(gla)} + 2  \mu^{(gla)} \ri
 \le
1- \frac{1}{2} \varsigma \ri
\end{array}
\ri \vspace{8pt}
\\
\*{\mathcal{D}}^{(2)} =
 \le
\begin{array}{lll}
\le   \mu^{(gla)}\ri \le 1- \frac{3}{2}
\varsigma \ri & 0 &
0 \\
0 &  \mu^{(gla)} \le 1 - \frac{4}{5}\varsigma \ri &  0
\\
0 & 0 &  \mu^{(gla)}
 \le
1- \frac{2}{3} \varsigma \ri
\end{array}
\ri
\end{array}
\right\},
\end{equation}
where the real--valued scalar $\varsigma$ controls the degree of
orthotropy. Thus, at fixed values of $\varsigma$ the component
materials may be viewed as being locally orthotropic. As in
\S\ref{iso_ell}, the densities of the component materials are taken
to be $\rho^{(1)} = \rho^{(ace)}$ and $\rho^{(2)} = \rho^{(gla)}$.
 The component materials are distributed as spherical
 particles (i.e., $a = b= c $).

\begin{figure}[h!]
\begin{center}
\resizebox{2.4in}{!}{\includegraphics{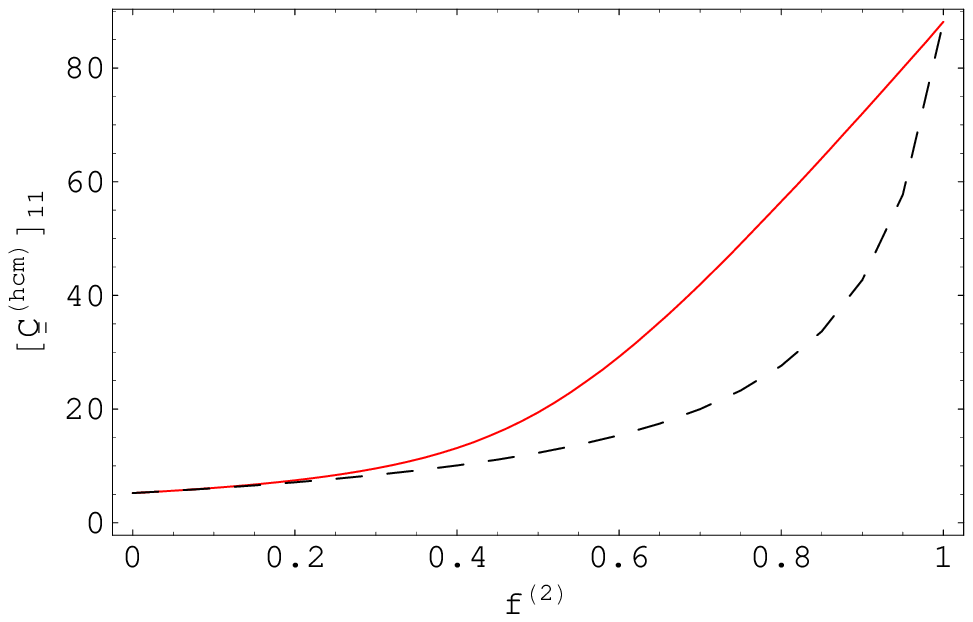}}
\resizebox{2.4in}{!}{\includegraphics{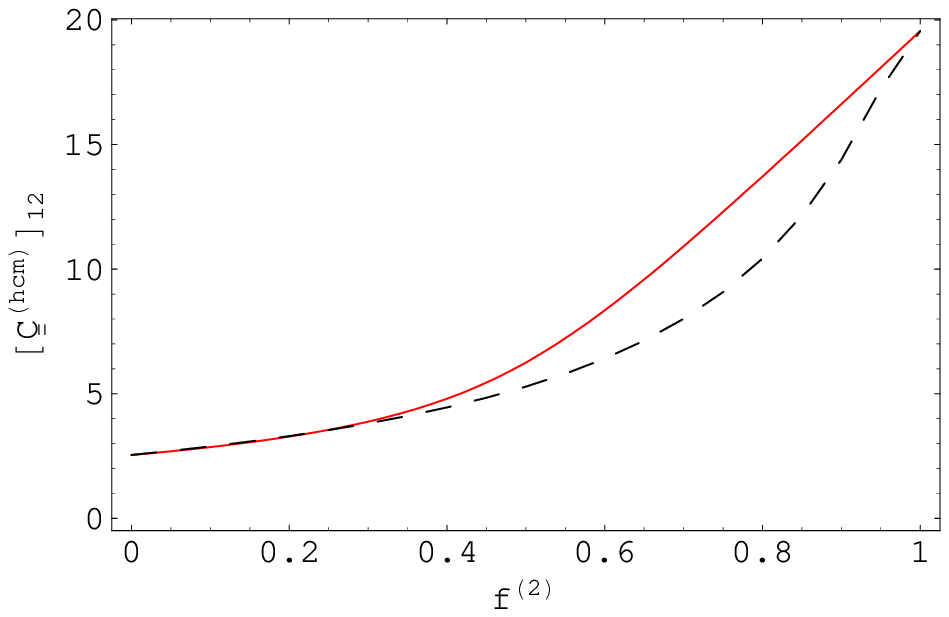}}
\resizebox{2.4in}{!}{\includegraphics{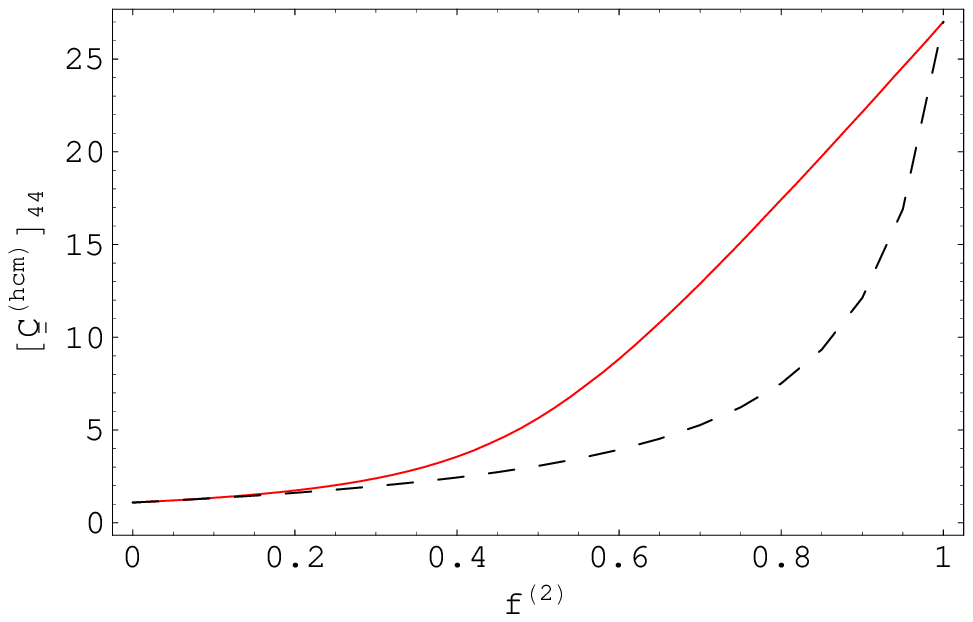}}
\caption{ Plots of $\les \, \*C^{(hcm)} \, \ris_{rs}$, with $ r s
\in \lec 11, 12, 33, 44 \ric$ (in GPa) as estimated using the
lowest--order SPFT (i.e., $hcm = ocm$) (red, solid curves) and the
Mori--Tanaka mean--field formalism (i.e., $hcm = MT$) (black, dashed
curves), against the volume fraction of component material `2'. The
component materials are distributed as spheres. Their constitutive
parameters are specified by \r{om1} and \r{om2}, with the orthotropy
parameter $\varsigma = 0.05$.} \label{glaaceplot_delta1}
\end{center}
\end{figure}

\begin{figure}[h!]
\begin{center}
\resizebox{2.4in}{!}{\includegraphics{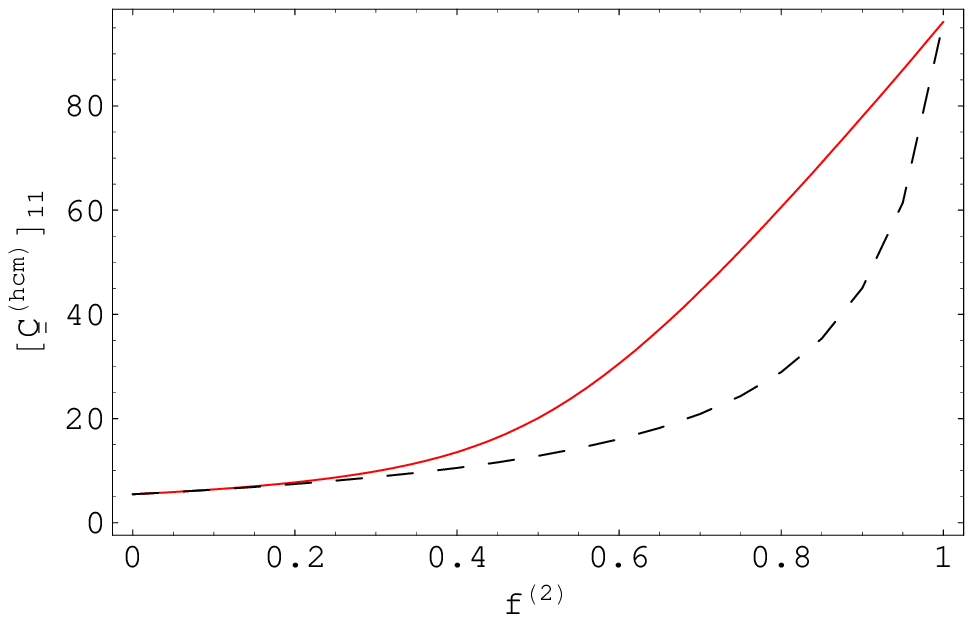}}
\resizebox{2.4in}{!}{\includegraphics{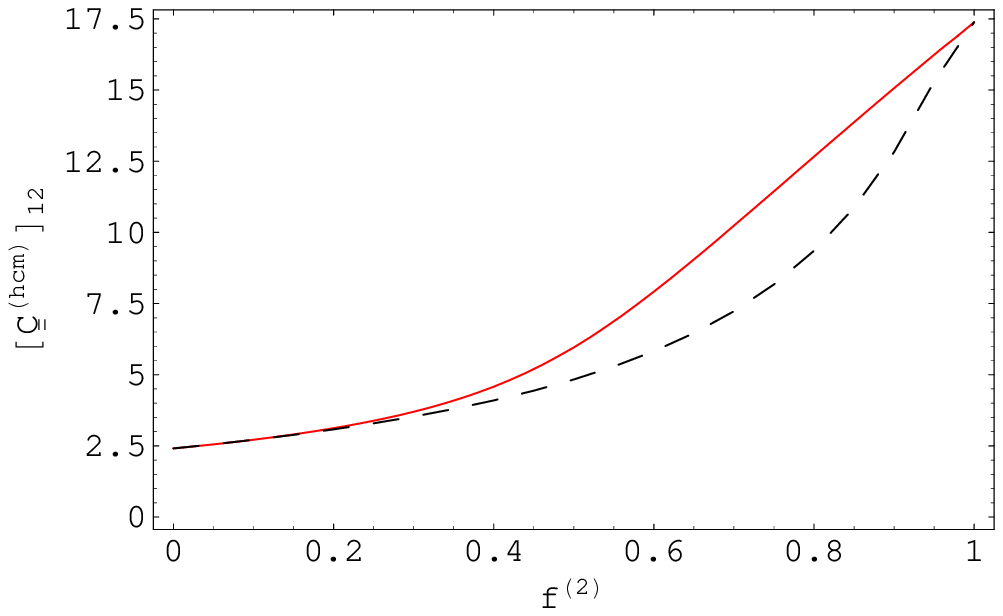}}
\resizebox{2.4in}{!}{\includegraphics{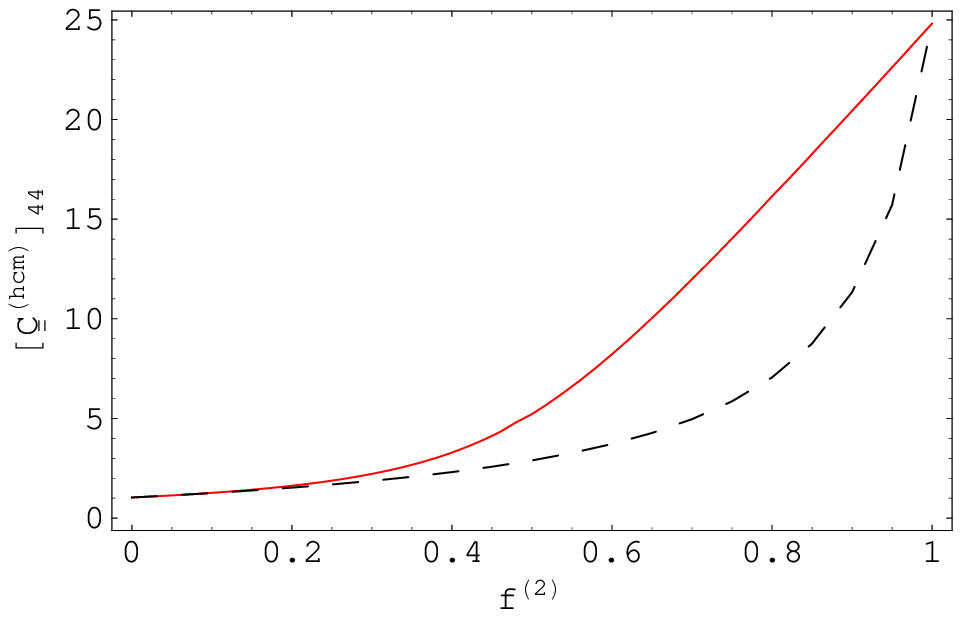}}
\caption{As Fig.~\ref{glaaceplot_delta1} but with orthotropy
parameter  $\varsigma=0.1$.}\label{glaaceplot_delta2}
\end{center}
\end{figure}

The  lowest--order SPFT and Mori--Tanaka estimates  for the  HCM
arising from orthotropic component materials characterized by
$\varsigma = 0.05$ and $\varsigma = 0.1$ are presented in
Fig.~\ref{glaaceplot_delta1} and \ref{glaaceplot_delta2},
respectively. As previously, only a representative selection of the
entries of $\*C^{(hcm)}$ are provided here. The plots for $\varsigma
= 0$, for which case the HCM is isotropic, are the ones displayed in
Fig.~\ref{glaaceplot_spher1}.
 In a
 manner resembling the scenario considered in
 \S\ref{iso_ell}, the lowest--order SPFT and the Mori--Tanaka
estimates are qualitatively similar, but the Mori--Tanaka estimates
display a greater deviation from the naive HCM estimate $f^{(1)}
\les \, \*C^{(1)} \, \ris_{pq} + f^{(2)} \les \, \*C^{(2)} \,
\ris_{pq}$ for mid--range values of $f^{(2)}$, at all values of
$\varsigma$.

\begin{figure}[h!]
\begin{center}
\resizebox{2.4in}{!}{\includegraphics{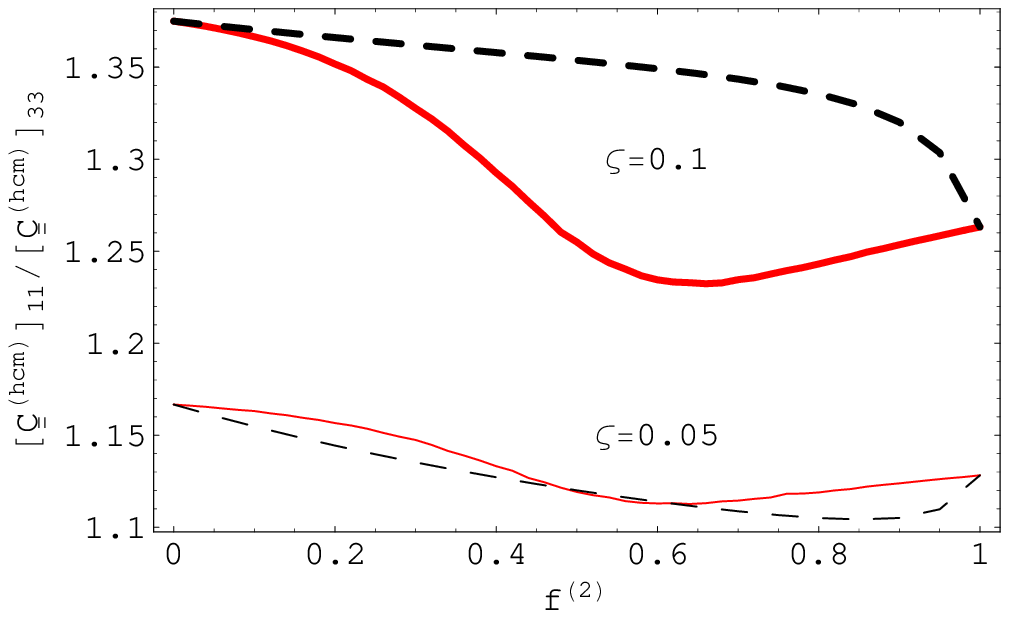}}
\resizebox{2.4in}{!}{\includegraphics{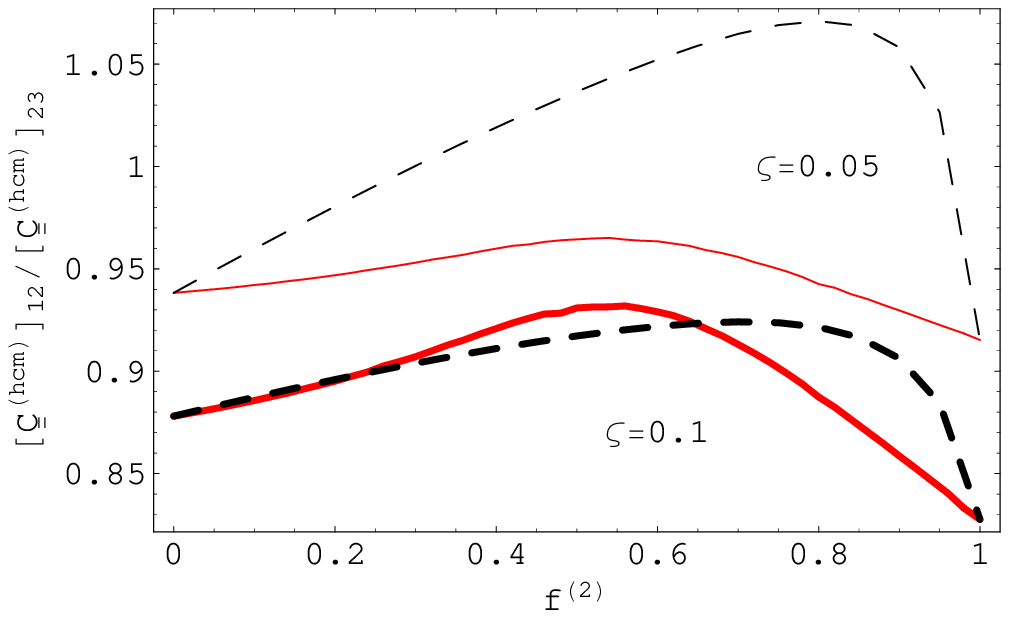}}
\resizebox{2.4in}{!}{\includegraphics{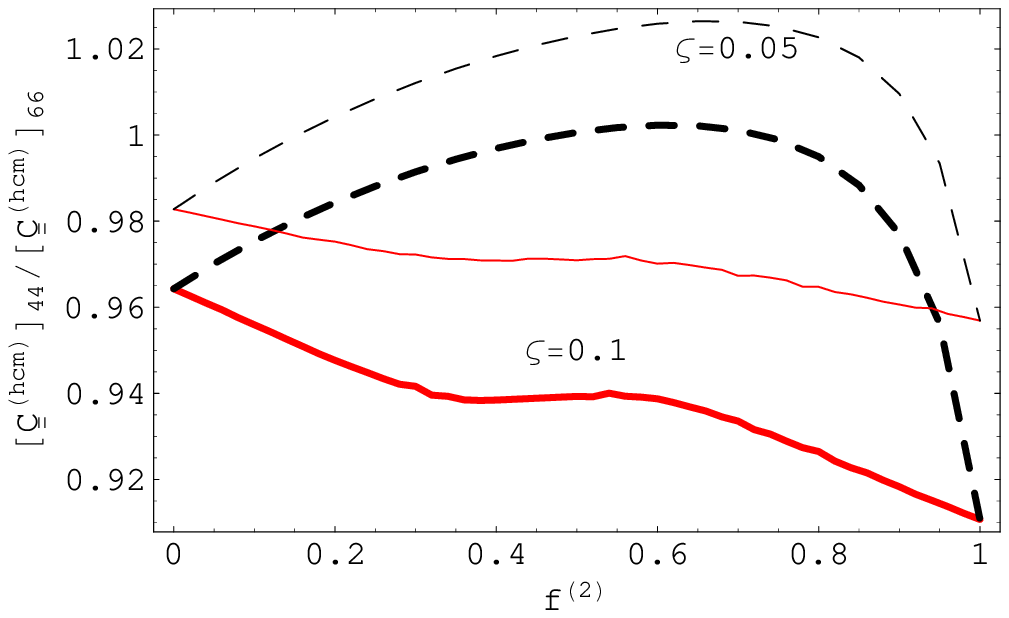}}
\caption{Plot of $\les \, \*C^{(hcm)} \, \ris_{11} / \les \,
\*C^{(hcm)} \, \ris_{33}$, $\les \, \*C^{(hcm)} \, \ris_{12} / \les
\, \*C^{(hcm)} \, \ris_{23}$ and $\les \, \*C^{(hcm)} \, \ris_{44} /
\les \, \*C^{(hcm)} \, \ris_{66}$ (in GPa) as estimated using the
lowest--order SPFT (i.e., $hcm = ocm$) (red, solid curves), the
Mori--Tanaka mean--field formalism (i.e., $hcm = MT$) (black, dashed
curves) against the volume fraction of component material `2'.
Component material `1' is acetal and component material `2' is
glass, as specified in \r{glass_acetal}. The component materials are
distributed as spheres with the orthotropy parameter
$\varsigma=0.05$ (thin curves) and $\varsigma=0.1$ (thick curves).
}\label{C11_C33_fig}
\end{center}
\end{figure}

The  degree of orthotropy exhibited by the HCM
 clearly increases as the value of $\varsigma$ increases, and
differences between the estimates of the lowest--order SPFT and the
Mori--Tanaka mean--field formalism also vary as $\varsigma$
increases. To explore this matter further, in Fig.~\ref{C11_C33_fig}
the associated ratios $\les \, \*C^{(hcm)} \, \ris_{11} / \les \,
\*C^{(hcm)} \, \ris_{33}$, $\les \, \*C^{(hcm)} \, \ris_{12} / \les
\, \*C^{(hcm)} \, \ris_{23}$ and $\les \, \*C^{(hcm)} \, \ris_{44} /
\les \, \*C^{(hcm)} \, \ris_{44}$ are plotted against $f^{(2)}$ for
$\varsigma = 0.05$ and $0.1$. The three different patterns are
portrayed in the three plots: for $\les \, \*C^{(hcm)} \, \ris_{11}
/ \les \, \*C^{(hcm)} \, \ris_{33}$ differences between the
lowest--order SPFT and the Mori--Tanaka estimates are larger for
when the HCM is more orthotropic; the reverse is the case for $\les
\, \*C^{(hcm)} \, \ris_{12} / \les \, \*C^{(hcm)} \, \ris_{23}$,
while for $\les \, \*C^{(hcm)} \, \ris_{44} / \les \, \*C^{(hcm)} \,
\ris_{66}$ there is no noticeable difference between the
lowest--order SPFT and Mori--Tanaka estimates as the degree of HCM
orthotropy is increased.

\begin{figure}[h!]
\begin{center}
\resizebox{2.4in}{!}{\includegraphics{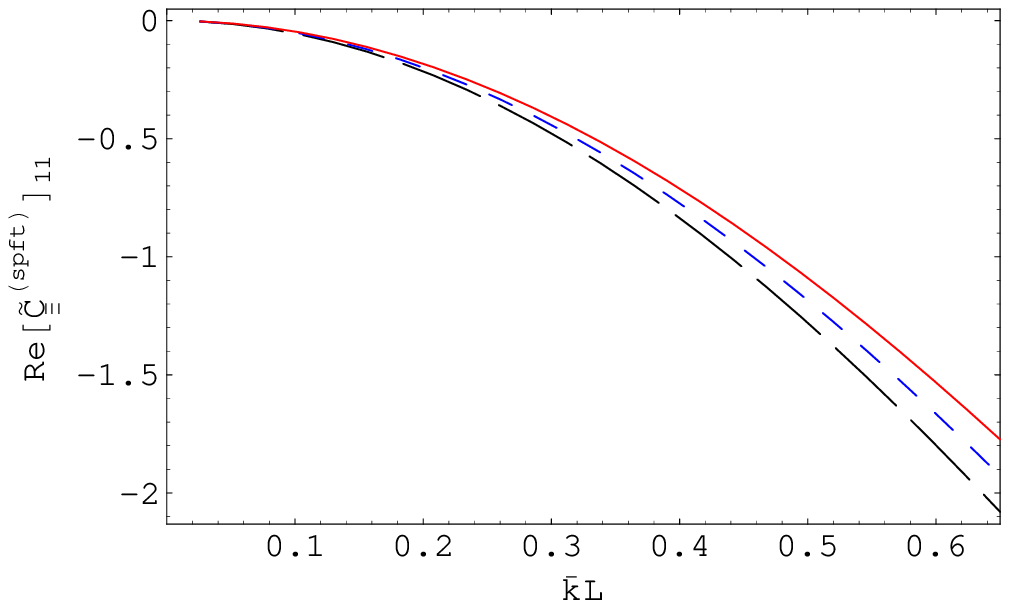}}
\resizebox{2.4in}{!}{\includegraphics{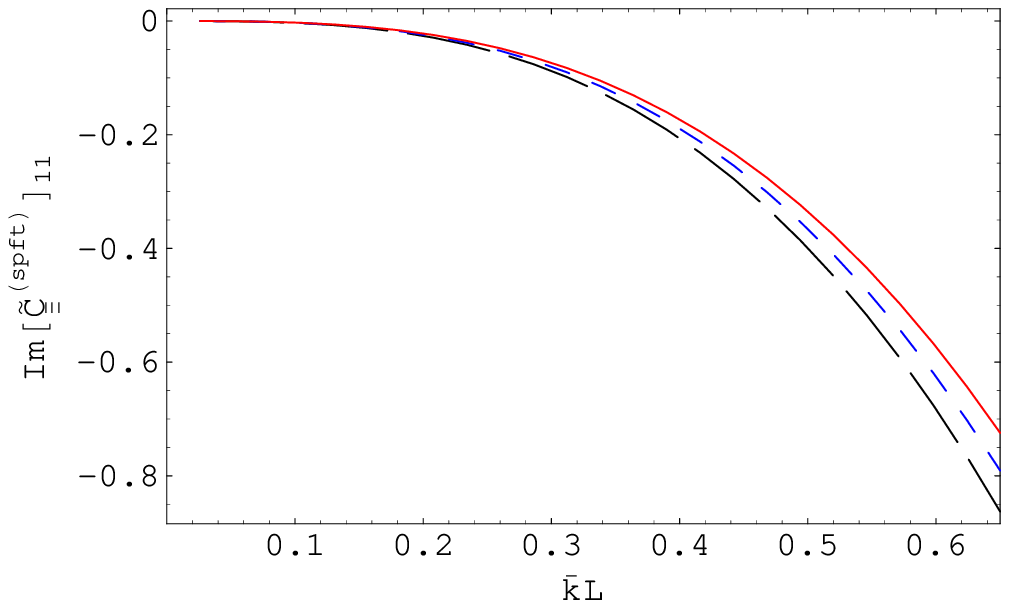}}
\resizebox{2.4in}{!}{\includegraphics{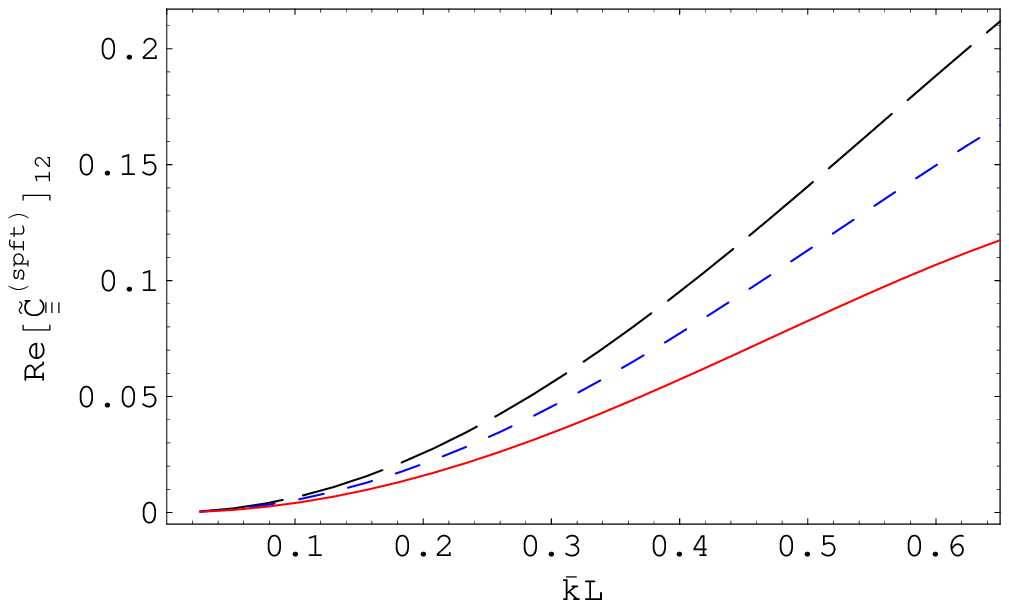}}
\resizebox{2.4in}{!}{\includegraphics{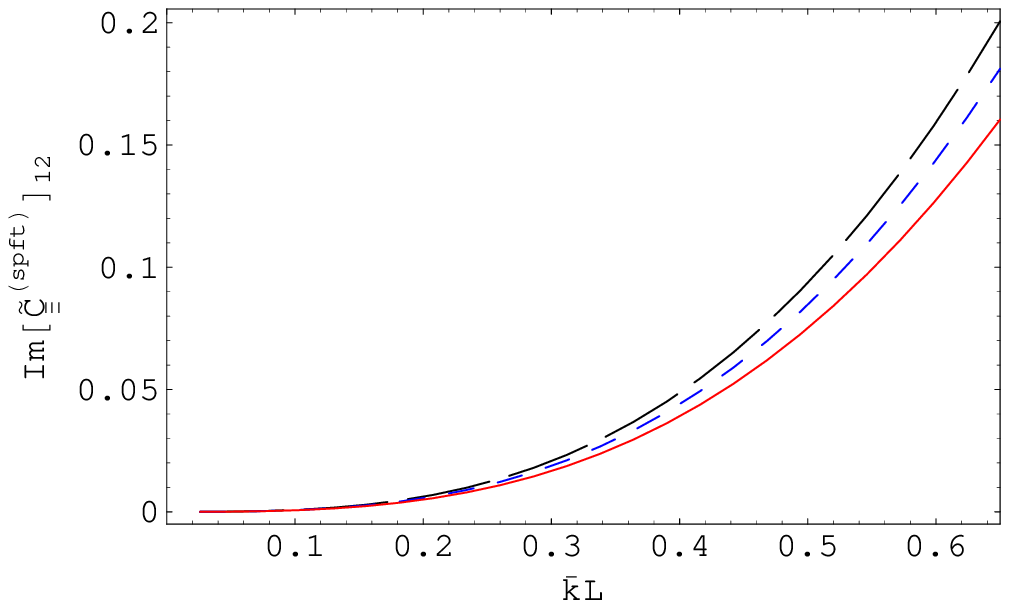}}
\resizebox{2.4in}{!}{\includegraphics{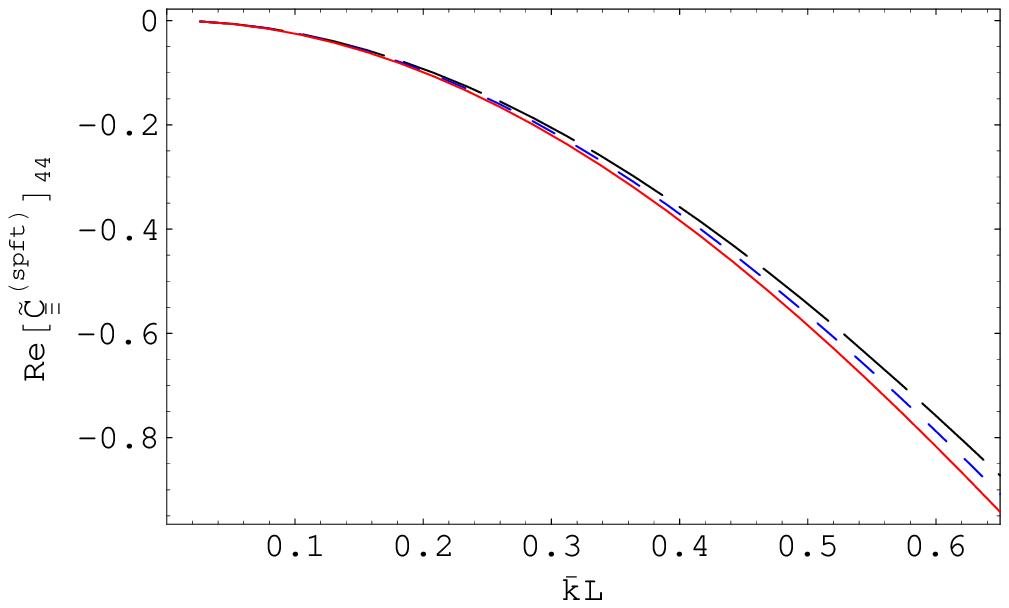}}
\resizebox{2.4in}{!}{\includegraphics{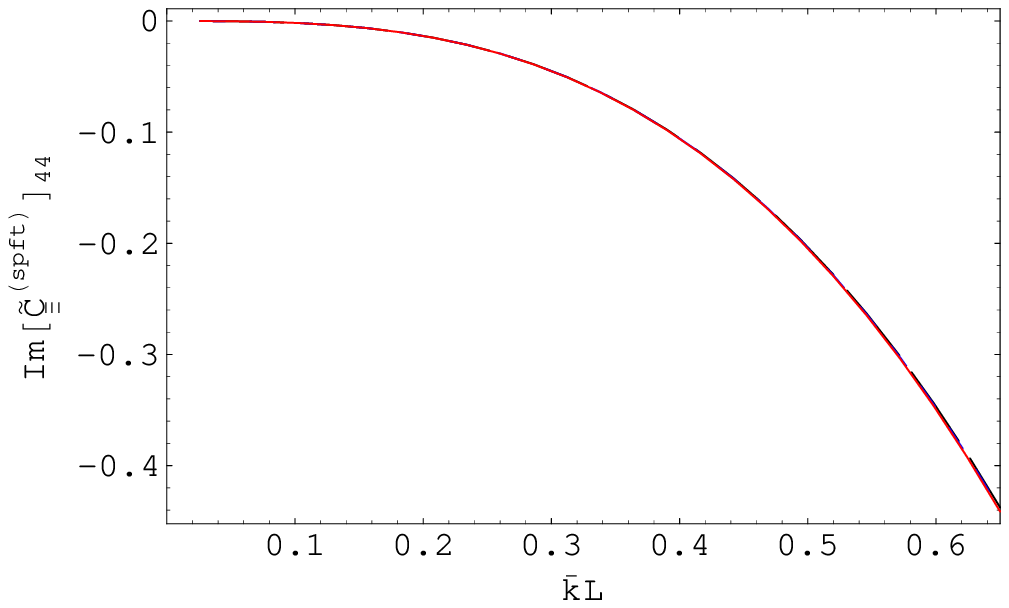}}
 \caption{The
real and imaginary parts of $\les \*{\tilde{C}}^{(spft)} \,
\ris_{rs}$, with $ rs \in \lec 11, 12, 44 \ric$ (in GPa) plotted as
functions of $\bar{k}L$, for $f^{(2)}=0.5$. The component materials
are distributed as spheres. Their constitutive parameters are
specified by \r{om1} and \r{om2}, with the orthotropy parameter
$\varsigma = 0$ (red, solid curves), $\varsigma =0.05$ (blue,
short-dashed curves) and $\varsigma =0.1$ (black, long-dashed
curves).}\label{Re_Cspftplot}
\end{center}
\end{figure}

Next we focus on the second--order SPFT estimate of the HCM
stiffness tensor. The real and imaginary parts of a representative
selection of entries of
$\*{\tilde{C}}^{(spft)}=\*C^{(spft)}-\*C^{(ocm)}$ are graphed
against $\bar{k} L$ in Fig.~\ref{Re_Cspftplot}. The volume fraction
is fixed at  $f^{(2)} = 0.5$. The values of the orthotropy parameter
$\varsigma$ are $0$, $0.05$ and $0.1$, in correspondence with the
calculations of Figs.~\ref{glaaceplot_spher1},
\ref{glaaceplot_delta1} and \ref{glaaceplot_delta2}. As we observed
in \S\ref{iso_ell}, the magnitude of the components of $
\*{\tilde{C}}^{(spft)} $ generally decrease as the HCM becomes more
orthotropic. Also, the second--order SPFT estimate $\*C^{(spft)}$
has components with nonzero imaginary parts, which implies that the
HCM is effectively dissipative even though the component materials
are nondissipative. Furthermore, the HCM becomes increasingly
dissipative as the correlation length increases, this effective
dissipation being attributable to scattering losses.

\begin{figure}[h!]
\begin{center}
\resizebox{2.4in}{!}{\includegraphics{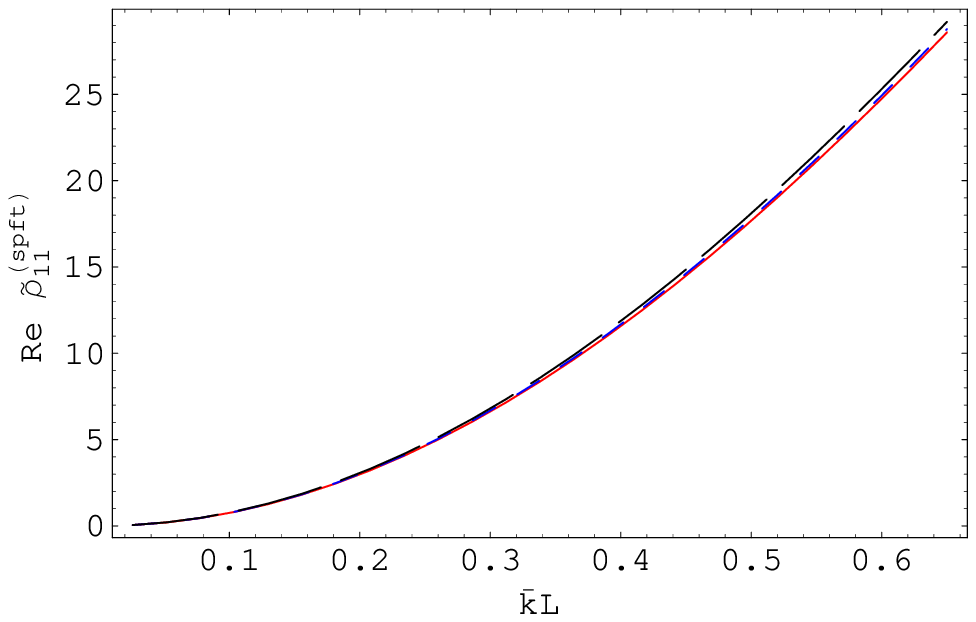}}
\resizebox{2.4in}{!}{\includegraphics{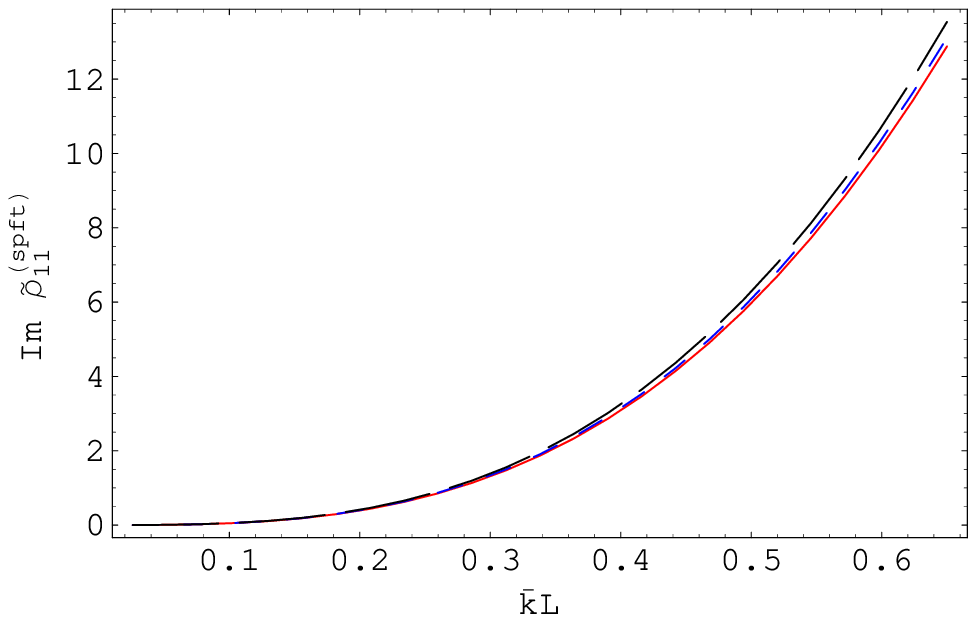}}
\resizebox{2.4in}{!}{\includegraphics{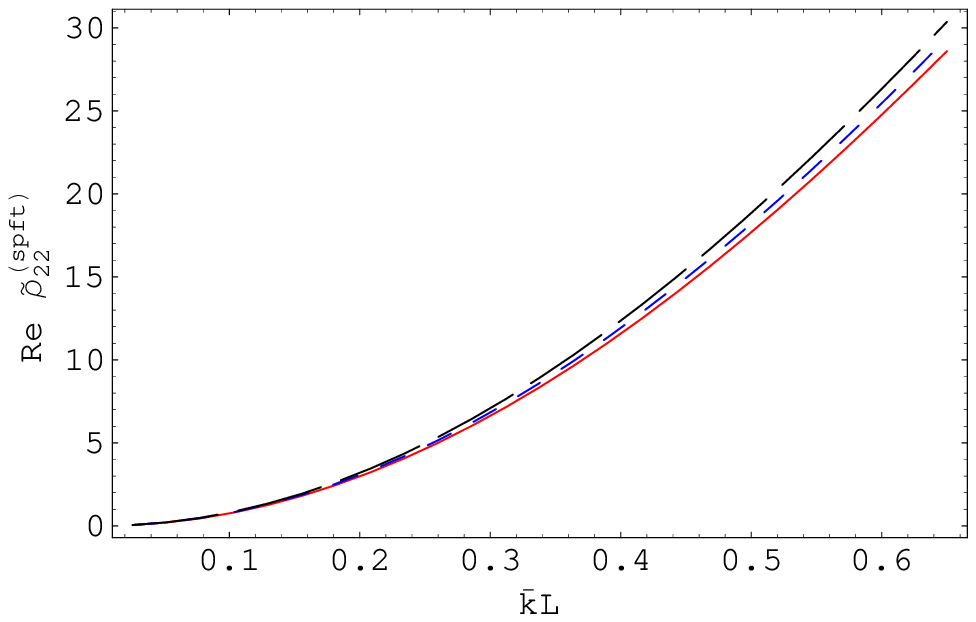}}
\resizebox{2.4in}{!}{\includegraphics{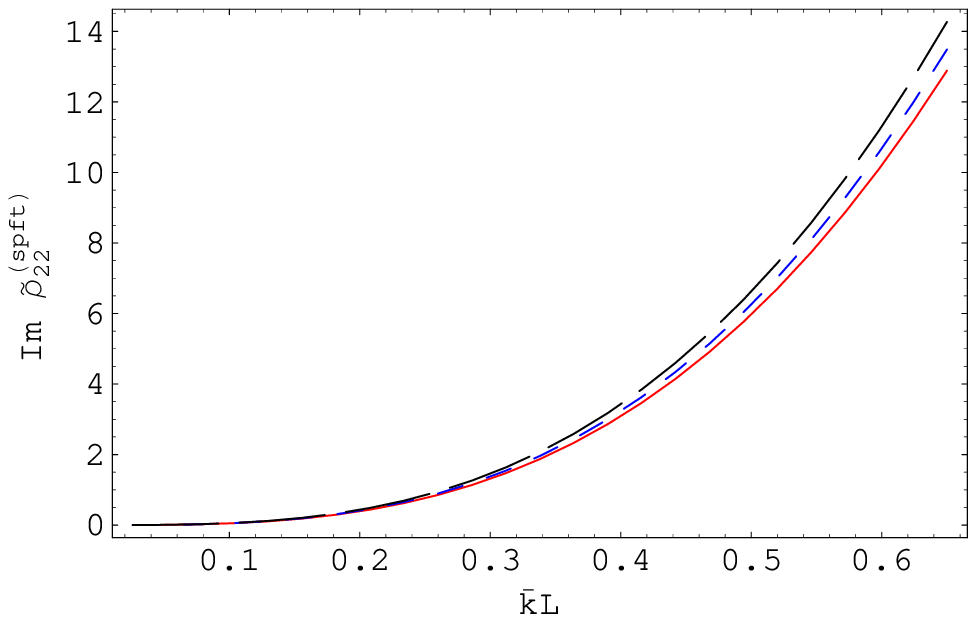}}
\resizebox{2.4in}{!}{\includegraphics{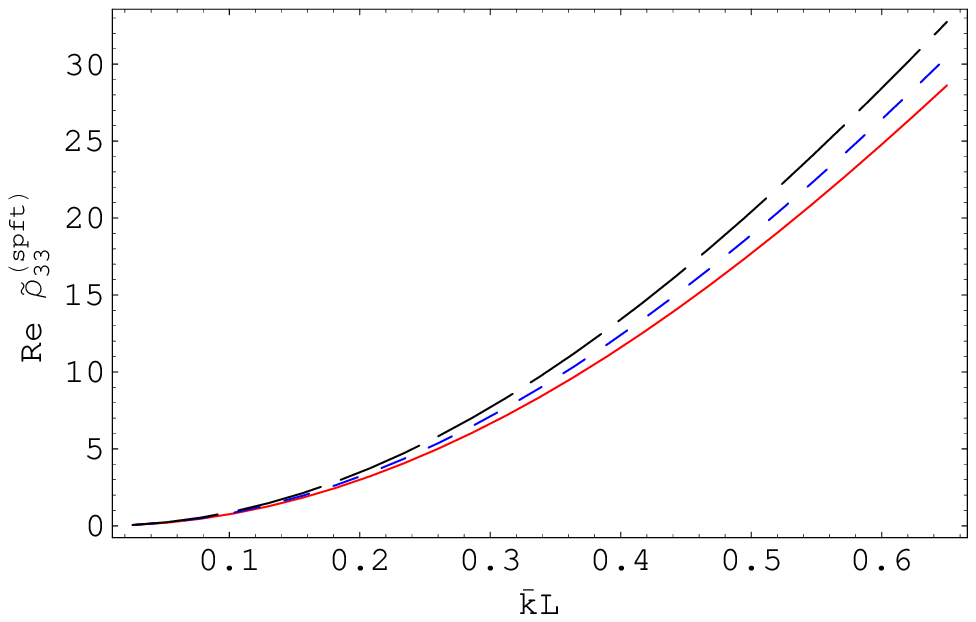}}
\resizebox{2.4in}{!}{\includegraphics{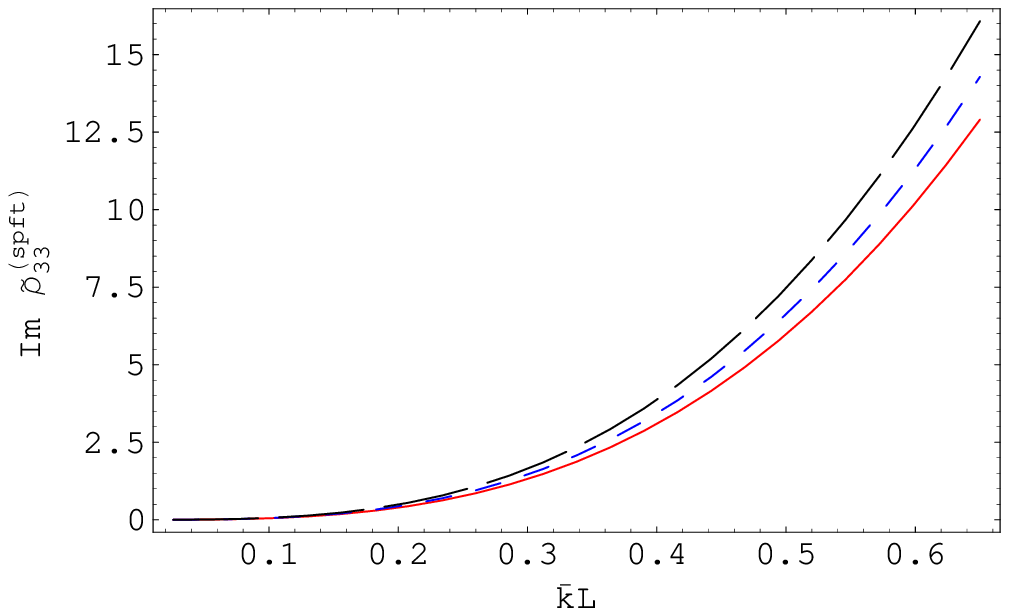}} \caption{ As
Fig.~\ref{Re_Cspftplot} but the quantities plotted are the real and
imaginary parts  of the excess of the second--order SPFT density
tensor over the density of the comparison material, i.e.,
$\tilde{\rho}^{(spft)}_{rr} =  \rho^{(spft)}_{rr} - \rho^{(ocm)}$,
($r \in \lec 1,2,3 \ric$), in $ \mbox{kg} \,
\mbox{m}^{-3}$.}\label{rhoplot}
\end{center}
\end{figure}

Finally,  the  real and imaginary parts of the second--order SPFT
density tensor $\tilde{\rho}^{(spft)}_{pq} = \rho^{(spft)}_{pq} -
\rho^{(ocm)}$ are plotted as functions of $\bar{k} L$ in
Fig.~\ref{rhoplot}. As previously in \S\ref{iso_ell}, the components
for $p \neq q$ are negligibly small so  only the $p=q$ components
are provided here. The density plots resemble those of the
corresponding stiffness tensor; i.e., the components
$\tilde{\rho}^{(spft)}_{pp}$ are much smaller than $\rho^{(ocm)}$
and they increase rapidly from zero as $L$ increases from zero. The
magnitudes of $\tilde{\rho}^{(spft)}_{pp}$ are smallest  when the
orthotropy parameter describing the component materials is  greatest.

\section{Closing remarks}

The elastodynamic SPFT has been  further developed, in order to
undertake numerical studies based on a specific choice of two--point
covariance function. From our theoretical considerations in
\S\ref{theory_section} and our representative numerical studies in
\S\ref{Num_Results}, involving generally orthotropic component
materials which are distributed as oriented ellipsoids, the
following conclusions were drawn:
\begin{itemize}

\item The lowest--order SPFT estimate of the HCM stiffness tensor
is qualitatively similar to that provided by the Mori--Tanaka
mean--field formalism.

\item Differences between the estimates of
the lowest--order SPFT and the Mori--Tanaka mean--field formalism
are greatest for mid--range values of the volume fraction.

\item Differences between the estimates of
the lowest--order SPFT and the Mori--Tanaka mean--field formalism
vary as the HCM becomes more orthotropic. The degree of orthotropy
of the HCM  may be increased by increasing either the degree of
orthotropy of component materials or
 the degree of eccentricity (nonsphericity) of the component particles.

\item The second--order SPFT provides a  low--frequency correction to the
quasi--static lowest--order estimates of the HCM stiffness tensor
and density. The correction vanishes as the correlation length tends
to zero.

\item The correction provided
by second--order SPFT, though  relatively small in magnitude,  is
highly significant as it indicates effective dissipation due to
scattering loss.

\item Differences between the second--order and
lowest--order SPFT estimates of the HCM constitutive parameters
diminish as the HCM becomes more orthotropic.

\end{itemize}

The ability to accommodate higher--order descriptions of the
distributional statistics of the component materials bodes well for
the future deployment of the SPFT in exploring the complex behaviour
of metamaterials as HCMs. Additionally, since the SPFT has been now
established for acoustic, electromagnetic and elastodynamic
homogenization scenarios, the prospect of considering HCMs with
mixed acoustic/elastodynamic/electromagnetic constitutive relations
beckons.

\vspace{15mm} \noindent {\bf Acknowledgements:} AJD is supported by
an \emph{Engineering and Physical Sciences Research Council} (UK)
 studentship. AL thanks the Charles Godfrey Binder Professorship Endowment
 for partial support.
\vspace{15mm}

\section*{Appendix A}

\subsection*{Matrix/tensor algebra}

A fourth--order tensor $A_{rstu}$ ($r,s,t,u \in \lec 1,2,3
\ric$) has $81$ components. If it obeys the symmetries
$A_{rstu} = A_{srtu} = A_{rsut} = A_{turs}$,
 it can be represented by a 9$\times$9 matrix $\*A $
with components $\les \,\*A \, \ris_{RS}$ ($R,S \in \lec 1,\ldots,9
\ric$) . Similarly, the nine entries of a second--order tensor
$B_{rs}$  ($r,s \in \lec 1,2,3 \ric$) may be expressed as a column
9--vector $\#B$ with components $\les \, \#B \, \ris_R$ ($R \in \lec 1,
\ldots, 9 \ric$). The scheme for converting between the tensor
subscript pairs $rs$ and $tu$ and the matrix indexes $RS$ or vector
index $R$ is provided in Table~\ref{MatrixConv}.

\begin{table}[h!]
  \centering
\begin{tabular}{|c|c||c|c||c|c|}
\hline
  $R,S$ & $rs,tu$ & $R,S$ & $rs,tu$ & $R,S$ & $rs,tu$ \\
  \hline
  \hline
  $1$  & $11$ & $4$ & $23$ or $32$ & $7$ & $23$ or $32$ \\
  $2$  & $22$ & $5$ & $13$ or $31$ & $8$ & $13$ or $31$ \\
  $3$  & $33$ & $6$ & $12$ or $21$ & $9$ & $12$ or $21$ \\
  \hline
\end{tabular}
  \caption{Conversion between tensor and matrix/vector subscripts.}\label{MatrixConv}
\end{table}

The most general 9$\times$9 matrix $\*A$ considered in this
paper has the form
\begin{equation} \l{A_form2}
\*A = \le \begin{array}{lll}  \*{\mbox{\boldmath$\alpha$}} & \*0 &
\*0
\\ \*0 & \*{\mbox{\boldmath$\beta$}}  & \*{\mbox{\boldmath$\beta$}}
\\
\*0 & \*{\mbox{\boldmath$\beta$}}  & \*{\mbox{\boldmath$\beta$}}
\end{array} \ri,
\end{equation}
where $ \*{\mbox{\boldmath$\alpha$}}$ is a general 3$\times$3
matrix, $\*{\mbox{\boldmath$\beta$}}$ is a  diagonal 3$\times$3
matrix,  and $\*0$ is the null 3$\times$3  matrix. If we define a
9$\times$9 matrix $\*A^\dagger$ as \c{Lakhjcm}
\begin{equation} \l{A_form22}
\*A^\dagger = \le \begin{array}{lll}
\*{\mbox{\boldmath$\alpha$}}^{-1} & \*0 & \*0 \vspace{4pt}
\\ \*0 &  \frac{1}{4} \, \*{\mbox{\boldmath$\beta$}}^{-1}  & \frac{1}{4} \, \*{\mbox{\boldmath$\beta$}}^{-1}
\vspace{4pt}
\\
\*0 & \frac{1}{4} \, \*{\mbox{\boldmath$\beta$}}^{-1}  & \frac{1}{4}
\, \*{\mbox{\boldmath$\beta$}}^{-1}
\end{array} \ri,
\end{equation}
then
 $\*A^{\dagger}\cdot \*A = \*A\cdot \*A^{\dagger}=
\*{\mbox{\boldmath$\tau$}}$, where $\*{\mbox{\boldmath$\tau$}}$ is
the 9$\times $9 matrix counterpart of the identity tensor
\begin{equation}
\tau_{rstu}=\frac{1}{2} \le
\delta_{rt}\delta_{su}+\delta_{ru}\delta_{st} \ri.
\end{equation}

\section*{Appendix B}

\subsection*{Eshelby matrix/tensor}

If  the component materials are orthotropic and distributed as
spherical particles (i.e., $a = b = c $), then  the tensor
counterpart of the 9$\times$9 Eshelby matrix is given as
 \cite{Numerical_Eshelby}
\begin{equation}\label{EshelbySint}
S^{(esh)}_{ijkl}=\frac{1}{8\pi}C^{(1)}_{mnkl}\int_{-1}^{+1}d\zeta_3
\int_{0}^{2\pi} d\omega \; \les \,
F_{imjn}(\overline{\vartheta})+F_{jmin}(\overline{\vartheta})\, \ris
,
\end{equation}
wherein
\begin{equation}
\left.
\begin{array}{l}
\displaystyle{
  F_{ijkl}(\overline{\vartheta}) = \frac{\overline{\vartheta}_k \overline{\vartheta}_l
  N_{ij}}{D}, \qquad
  N_{ij}(\overline{\vartheta}) = \frac{1}{2}\epsilon_{ikl}\epsilon_{jmn} K_{km}K_{ln}} \vspace{4pt}\\
  \displaystyle{ D = \epsilon_{mnl}K_{m1}K_{n2}K_{l3}, \qquad
  K_{ik}= C^{(1)}_{ijkl} \overline{\vartheta}_j \overline{\vartheta}_l} \qquad \\
\displaystyle{  \overline{\vartheta}_1 = \frac{\zeta_1}{a}, \qquad
 \overline{\vartheta}_2 = \frac{\zeta_2}{b}, \qquad \overline{\vartheta}_3 = \frac{\zeta_3}{c}} \vspace{4pt}
  \\
 \displaystyle{ \zeta_1 = (1-\zeta_3^2)^{1/2}\cos(\omega), \qquad
  \zeta_2=(1-\zeta_3^2)^{1/2}\sin(\omega), \qquad
  \zeta_3=\zeta_3}
  \end{array} \right\},
\end{equation}
with  $\epsilon_{ijk}$ being the Levi--Civita symbol. The integrals
in \r{EshelbySint} can be evaluated using standard numerical methods
\c{num_methods}.

If the component materials are isotropic and distributed as
ellipsoidal particles described by the shape matrix $\*U$, then the
 Eshelby matrix has the form represented in  \r{A_form2} with distinct components
 given as \cite{Mura_book}
\begin{eqnarray}
\les \,  \*S^{(esh)} \,\ris_{11} &=& \frac{3a^2I_{\alpha \alpha}+\le
1-2\nu^{(1)} \ri I_\alpha}{8\pi\le 1-\nu^{(1)} \ri}, \end{eqnarray}
\begin{eqnarray} \les \,
\*S^{(esh)}  \,\ris_{12} &=& \frac{3b^2I_{\alpha \beta}- \le
1-2\nu^{(1)} \ri I_\alpha}{8\pi \le 1-\nu^{(1)} \ri}, \end{eqnarray}
\begin{eqnarray} \les \,
\*S^{(esh)} \,\ris_{13} &=& \frac{3c^2I_{\alpha \gamma}-\le
1-2\nu^{(1)} \ri I_\alpha}{8\pi\le 1-\nu^{(1)} \ri}, \end{eqnarray}
\begin{eqnarray} \les \,
\*S^{(esh)} \,\ris_{21} &=& \frac{3a^2I_{\alpha \beta }-\le
1-2\nu^{(1)} \ri I_\beta}{8\pi \le 1-\nu^{(1)} \ri} ,\end{eqnarray}
\begin{eqnarray} \les \,
\*S^{(esh)} \,\ris_{22} &=& \frac{3b^2I_{\beta \beta}+\le
1-2\nu^{(1)} \ri I_\beta}{8\pi \le 1-\nu^{(1)} \ri }, \end{eqnarray}
\begin{eqnarray} \les \,
\*S^{(esh)} \,\ris_{23} &=& \frac{3c^2I_{\beta \gamma}+ \le
1-2\nu^{(1)} \ri I_\beta}{8\pi \le 1-\nu^{(1)} \ri },
\end{eqnarray}
\begin{eqnarray}
\les \,  \*S^{(esh)} \,\ris_{31} &=& \frac{3a^2I_{\alpha \gamma }-
\le 1-2\nu^{(1)} \ri I_\gamma}{8\pi \le 1-\nu^{(1)} \ri },
\end{eqnarray}
\begin{eqnarray} \les \,
\*S^{(esh)} \,\ris_{32} &=& \frac{3b^2I_{\beta \gamma}-\le
1-2\nu^{(1)} \ri I_\gamma}{8\pi \le 1-\nu^{(1)} \ri },
\end{eqnarray}
\begin{eqnarray} \les \,
\*S^{(esh)} \,\ris_{33} &=& \frac{3c^2I_{\gamma \gamma}+ \le
1-2\nu^{(1)} \ri I_\gamma}{8\pi \le 1-\nu^{(1)} \ri } ,
\end{eqnarray}
\begin{eqnarray}
 \les \,
\*S^{(esh)} \,\ris_{44} &=& \frac{3(b^2+c^2)I_{\beta \gamma}+ \le 1
-2\nu^{(1)} \ri (I_\beta +I_\gamma)}{16\pi \le 1-\nu^{(1)} \ri}
,\end{eqnarray}
\begin{eqnarray}
\les \,  \*S^{(esh)} \,\ris_{55} &=& \frac{3(a^2+c^2)I_{\alpha
\gamma}+\le 1-2\nu^{(1)} \ri (I_\alpha +I_\gamma)}{16\pi \le
1-\nu^{(1)} \ri} ,\end{eqnarray}
\begin{eqnarray} \les \,  \*S^{(esh)} \,\ris_{66} &=&
  \frac{3(a^2+b^2)I_{\alpha \beta}+\le 1-2\nu^{(1)} \ri(I_\alpha+I_\beta)}{16\pi \le 1-\nu^{(1)} \ri},
\end{eqnarray}
where $\nu^{(1)} = \displaystyle{\frac{\lam^{(1)}}{2\le
\lam^{(1)}+\mu^{(1)}\ri}}$ is the Poisson ratio of component material
`1'. For the case $a>b>c$ we have
\begin{equation}
\left. \begin{array}{l}
  I_{\alpha} = \displaystyle{\frac{4\pi
  abc}{(a^2-b^2)(a^2-c^2)^{1/2}}\les
  \, F(\tilde{\theta},\tilde{k})-E(\tilde{\theta},\tilde{k})\ris}
   \vspace{4pt} \\
  I_{\gamma} = \displaystyle{\frac{4\pi
  abc}{(b^2-c^2)(a^2-c^2)^{1/2}}\les
  \, \frac{b}{ac}(a^2-c^2)^{1/2}-E(\tilde{\theta},\tilde{k})\ris }
   \vspace{4pt} \\
  I_{\beta} = \displaystyle{4\pi-(I_\alpha +I_\gamma)}, \qquad
  I_{\alpha \beta} = \displaystyle{\frac{I_\alpha -I_\beta}{3(b^2-a^2)}}, \qquad
  I_{\alpha \gamma} = \displaystyle{\frac{I_\alpha -I_\gamma}{3(c^2-a^2)}} , \qquad
  I_{\beta \gamma} = \displaystyle{\frac{I_\beta -I_\gamma}{3(c^2-b^2)}}  \vspace{4pt} \\
  I_{\alpha \alpha} = \displaystyle{\frac{4\pi}{3a^2}-(I_{\alpha \beta}+I_{\alpha \beta})}, \qquad
  I_{\beta \beta} = \frac{4\pi}{3b^2}-(I_{\alpha \beta}+I_{\beta \gamma}), \qquad
  I_{\gamma \gamma} = \frac{4\pi}{3c^2}-(I_{\alpha \gamma}+I_{\beta \gamma}) \end{array} \right\},
\end{equation}
with the elliptic integrals given by
\begin{equation}
\left.
\begin{array}{l}
\displaystyle{E(\tilde{\theta},\tilde{k}) = \int_0^{\tilde{\theta}}
d\phi \;(1-\tilde{k}^2\sin^2\phi)^{1/2}} \vspace{4pt}
\\
\displaystyle{F(\tilde{\theta},\tilde{k}) = \int_0^{\tilde{\theta}}
d\phi (1-\tilde{k}^2\sin^2\phi)^{-1/2}} \end{array} \right\},
\end{equation}
wherein
\begin{equation}
\tilde{\theta}=\sin^{-1}\frac{(a^2-c^2)^{1/2}}{a}, \qquad
\tilde{k}=\frac{(a^2-b^2)^{1/2}}{(a^2-c^2)^{1/2}}.
\end{equation}

\newpage

\newpage

\newpage

\newpage

\newpage

\newpage

\newpage

\newpage

\newpage

\newpage

\begin{thebibliography}{99}

\bibitem{Akhlesh_book}
Lakhtakia, A. (Ed.), (1996)  \emph{Selected Papers on Linear Optical
Composite Materials}. Bellingham, WA, USA: SPIE.


\bibitem{Milton_book}
 Milton, G.W. (2002) \emph{The Theory of Composites}.
  Cambridge, UK: Cambridge University Press.

\bibitem{Walser}
  Walser, R.M. (2003) Metamaterials: an introduction.
\emph{Introduction to Complex Mediums for Optics and
Electromagnetics} (W.S. Weiglhofer, A. Lakhtakia eds.).
 Bellingham, WA, USA: SPIE,  pp. 295--316.

\bibitem{Mei}
Mei, J., Liu, Z., Wen, W. \& Sheng, P. (2007) Effective dynamic mass
density of composites. \emph{Phys. Rev. B}, {\bf 76}, 134205.


\bibitem{Lakes_01}
Lakes, R.S. (2001) Extreme damping in composite materials with a
negative stiffness phase. \emph{Phys. Rev. Lett.}, {\bf 86},
2897--2900.

\bibitem{Fang}
Fang, N., Xi, D., Xu, J., Ambati, M., Srituravanich, W., Sun, C. \&
Zhang Y. (2006) Ultrasonic metamaterials with negative modulus.
\emph{Nature Materials}, {\bf 5}, 452--456.



\bibitem{Rama}
Ramakrishna, S.A. (2005) Physics of negative refractive index
materials. \emph{Rep. Prog. Phys.}, {\bf 68},  449--521.


\bibitem{Ryzhov}
 Ryzhov, Yu A.  \&  Tamoikin, V.V. (1970)
 Radiation and propagation of
electromagnetic waves in randomly inhomogeneous media. \emph{
Radiophys. Quantum Electron.}, {\bf 14}, 228--233.



\bibitem{Hill_1963}
Hill R. (1963) A self--consistent mechanics of composite materials.
\emph{J. Mech. Phys. Solids}, \textbf{13}, 213--222.


\bibitem{Hill_1965}
Hill R. (1965) Elastic properties of reinforced solids: some
theoretical principles. \emph{J. Mech. Phys. Solids}, \textbf{11},
357--372.

\bibitem{Bud}
Budiansky B. (1965) On the elastic moduli of some heterogeneous
materials. \emph{J. Mech. Phys. Solids}, \textbf{13}, 223--227.

\bibitem{Sabina_Willis}
Sabina, F.J. \& Willis, J.R. (1988) A simple self-consistent
analysis of wave propagation in particulate composites. \emph{Wave
Motion}, \textbf{10}, 127--142.

\bibitem{Kim}
Kim, J.--Y. (2004) On the generalized self--consistent model for
elastic wave propagation in composite materials. \emph{Int. J.
Solids Structures}, {\bf 41}, 4349--4360.

\bibitem{Kanaun}
Kanaun, S.K. \& Levin, V.M. (2005) Propagation of shear elastic
waves in composites with a random set of spherical inclusions
(effective field approach). \emph{Int. J. Solids Structures}, {\bf
42}, 3971--3997.

\bibitem{Wang_Qin}
Wang, Y. \& Qin, Q.--H. (2007) A generalized self consistent model
for effective elastic moduli of human dentine. \emph{Compos. Sci.
Technol.}, {\bf 67}, 1553--1560.

\bibitem{Twersky}
Twersky, V. (1962) On scattering of waves by random distributions.
I. Free--space scatterer formalism. \emph{J. Math. Phys.}, {\bf 3},
700--715.

\bibitem{Linton}
Linton, C.M.  \& Martin, P.A.  (2006) Multiple scattering by
multiple spheres: a new proof of the Lloyd--Berry formula for the
effective wavenumber. \emph{SIAM J. Appl. Math.}, {\bf 66},
1649--1668.


\bibitem{Avila}
\'{A}vila--Carrera, R., S\'{a}nchez--Sesma, F.J. \& Avil\'{e}s, J.
(2008) Transient response and multiple scattering of elastic waves
by a linear array of regulaly distributed cylindrical obstacles:
Anti--plane S--wave analytical solution. \emph{Geof\'{i}s. Int.},
{\bf 47}, 115--126.

\bibitem{Datta}
Datta, S.K. \& Ledbetter, H.M. (1986) Effective wave speeds in an
SiC--particle--reinforced Al composite. \emph{J. Acoust. Soc. Am.},
{\bf 79}, 239--248.

\bibitem{Varadan}
Varadan, V.K., Ma, Y. \& Varadan, V.V. (1989) Scattering and
attenuation of elastic waves in random media. \emph{Pure Appl.
Geophys.}, {\bf 131}, 577--603.

\bibitem{Maurel}
Maurel, A., Mercier, J.--F. \& Lund, F. (2004) Elastic wave
propagation through a random array of dislocations. \emph{Phys. Rev.
B}, {\bf  70}, 024303.

\bibitem{HS_1962}
Hashin, Z. \& Shtrikman, S. (1962) On some variational principles in
anisotropic and nonhomogeneous elasticity. \emph{J. Mech. Phys.
Solids}, \textbf{10}, 335--342.

\bibitem{Kroner}
Kr\"oner, E. (1977) Bounds for the effective elastic moduli of
disordered materials. \emph{J. Mech. Phys. Solids}, \textbf{25},
137--155.

\bibitem{Willis}
Willis, J.R. (1981) Variational and related methods for the overall
properties of composites. \emph{Advances in Applied Mechanics} (ed.
C.S. Jeh ed.), vol. 21. New York, NY, USA: Academic, pp.1--79.

\bibitem{Talbot_Willis}
Talbot, D.R.S. \& Willis, J.R. (1982) Variational estimates for the
dispersion and attenuation of waves in random composites I. General
theory. \emph{Int. J. Solids Structures}, \textbf{18}, 673--683.

\bibitem{Hashin_83}
Hashin, Z. (1983) Analysis of composite materials~---~a survey.
\emph{J. Appl. Mech.}, \textbf{50}, 481--505.


\bibitem{TK81}
 Tsang, L.  \&    Kong, J.A. (1981)
 Scattering of electromagnetic waves
from random media with strong permittivity fluctuations. \emph{
Radio Sci.},
 {\bf 16},  303--320.

\bibitem{Genchev}
 Genchev, Z.D. (1992)
 Anisotropic and gyrotropic version of Polder and
van Santen's mixing formula. \emph{ Waves Random Media}, {\bf 2},
99--110.

\bibitem{ML95}
 Michel, B.  \&  Lakhtakia, A. (1995)
 Strong--property--fluctuation theory
for homogenizing chiral particulate composites. \emph{ Phys. Rev.
E}, {\bf 51}, 5701--5707.


\bibitem{spft_form}
Mackay, T.G., Lakhtakia, A. \& Weiglhofer, W.S. (2000)
Strong-property-fluctuation theory for homgenization of
bianisotropic composites: Formulation. \emph{Phys. Rev. E},
\textbf{62}, 6052--6064. Erratum: (2001) {\bf 63}, 049901.



\bibitem{Cui}
Cui, J. \&  Mackay, T.G.  (2007) Depolarization regions of nonzero
volume in bianisotropic homogenized composites. \emph{Waves Random
Complex Media},  {\bf 17}, 269--281.


\bibitem{Zhuck_acoustics}
Zhuck, N.P. (1996) Strong fluctuation theory for a mean acoustic
field in a random fluid medium with statistically anisotropic
perturbations. \emph{J. Acoust. Soc. Am.}, {\bf 99}, 46--54.


\bibitem{spft_zhuck1}
Zhuck, N.P. \& Lakhtakia, A. (1999) Effective constitutive
properties of a disordered elastic solid medium via the
strong-fluctuation approach. \emph{Proc R. Soc. Lond. A},
\textbf{455}, 543--566.


\bibitem{Mori-Tanaka}
Mori, T. \& Tanaka, K. (1993) Average stress in matrix and average
elastic energy of materials misfitting inclusions. \emph{Acta
Metallurgica}, \textbf{21}, 571--574.

\bibitem{Benveniste}
Benveniste, Y. (1987) A new approach to the application of
Mori-Tanaka's theory in composite materials. \emph{Mech. Materials}
{\bf 6}, 147--157.

\bibitem{Lakhjcm}
Lakhtakia, A. (2002) Microscopic model for elastostatic and
elastodynamic excitation of chiral sculptured thin films. \emph{J.
Compos. Mater.}, \textbf{36}, 1277--1298.


\bibitem{Akhlesh_book_intro}
Lakhtakia, A.  (1996) Introduction  \emph{Selected Papers on Linear
Optical Composite Materials} (A. Lakhtakia ed.). Bellingham, WA,
USA: SPIE.


\bibitem{M08}
Mackay, T.G. (2008) Lewin's homogenization formula revisited for
nanocomposite materials.
  \emph{J. Nanophoton.}, {\bf 2},  029503.

\bibitem{Cerveny_2006}
\v{C}erven\'y, V. \& P\v{s}en\v{c}ik, I. (2006) Energy flux in
viscolelastic anisotropic media. \emph{Geophys. J. Int.},
\textbf{166}, 1299--1317.


\bibitem{Ting}
Ting, T.C.T (1996) \emph{Anisotropic Elasticity}. New York, NY, USA:
Oxford University Press.


\bibitem{TKN82}
 Tsang, L,   Kong, J.A. \&  Newton, R.W. (1982)
 Application of strong
fluctuation random medium theory to scattering of electromagnetic
waves from a half--space of dielectric mixture.  \emph{IEEE Trans.
Antennas Propagat.}, {\bf 30}, 292--302.

\bibitem{MLW01b}
  Mackay, T.G.,  Lakhtakia, A.  \&   Weiglhofer, W.S. (2001)
Homogenisation of similarly oriented, metallic, ellipsoidal
inclusions using the bilocally approximated
strong--property--fluctuation theory. \emph{ Opt. Commun.}, {\bf
107}, 89--95.


\bibitem{MNLS}
Narasimhan, M.N.L. (1993) \emph{Principles of Continuum Mechanics}.
New York, NY, USA: Wiley.

\bibitem{Parnell}
Parnell, W.J. \& Abrahams, I.D. (2008) Homogenization for wave
propagation in periodic fibre--reinforced media with complex
microstructure. I--Theory. \emph{J. Mech. Phys. Solids}, {\bf 56},
2521--2540.

\bibitem{Bagnara}
Bagnara R. (1995) A unified proof for the convergence of Jacobi and
Gauss--Seidel methods. \emph{SIAM Review}, \textbf{37}, 93--97.

\bibitem{Kwok}
Kwok, Y.K. (2002) \emph{Applied Complex Variables for Scientists and
Engineers}. Cambridge, UK: Cambridge University Press.


\bibitem{num_methods}
Press, W.H.,    Flannery, B.P.,    Teukolsky, S.A.   \&
 Vetterling,
W.T. (1992)
 \emph{Numerical Recipes in Fortran}, 2nd. edition.
 Cambridge, UK: Cambridge University Press.

\bibitem{Biwa}
Biwa, S., Kobayashi, F. \& Ohno, N. (2007) Influence of disordered
fiber arrangement on SH wave transmission in unidirectional
composites. \emph{Mech. Mater.}, {\bf 39}, 1--10.

\bibitem{Willis85}
Willis, J.R. (1985) The nonlocal influence of density variations in
a composite. \emph{Int. J. Solids Structures}, {\bf 21}, 805--817.

\bibitem{Milton_NJP2007}
Milton, G.W. (2007) New metamaterials with macroscopic behavior
outside that of continuum elastodynamics. \emph{New J. Physics},
{\bf 9}, 359.


\bibitem{Mura_book}
Mura, T. (1987) \emph{Micromechanics of Defects in Solids}.
Dordrecht, The Netherlands: Martinus
 Nijhoff Publishers.


\bibitem{Ferrari}
Ferrari, M. \& Filipponi, M. (1991) Appraisal of current
homogenizing techniques for the elastic response of porous and
reinforced glass. \emph{J. Am. Ceramic Soc.}, {\bf 74}, 229--231.

\bibitem{Hu_Weng}
Hu, G.K. \& Weng, G.J. (2000) Some reflections on the Mori--Tanaka
and Ponte Casta\~{n}eda--Willis methods with randomly oriented
ellipsoidal inclusions. \emph{Acta Mechanica}, {\bf 140}, 31--40.

\bibitem{Mercier}
Mercier, S. \& Molinari, A. (2008) Homogenization of
elastic--viscoelastic heterogeneous materials: Self--consistent and
Mori--Tanaka schemes. \emph{Int. J. Plasticity}, (in press).
[doi:10.1016/j.ijplas.2008.08.006]

\bibitem{Eshelby}
Eshelby, J.D. (1957) The determination of the elastic field of an
ellipsoidal inclusion, and related problems. \emph{Proc. R. Soc.
Lond. A}, {\bf 241}, 376--396.


\bibitem{MacMillan}
James, A.M. \& Lord, M.P. (1992) \emph{MacMillan's Chemical and
Physical Data}. London, UK: MacMillan Press.

\bibitem{Shackelford}
Shackelford, J.F. (2005) \emph{Introduction to Materials Science for
Engineers}, 6th. edition. Upper Saddle River,  NJ, USA: Pearson
Prentice Hall.

\bibitem{Numerical_Eshelby}
Gavazzi, A.C. \& Lagoudas, D.C. (1990) On the numerical evaluation
of Eshelby's tensor and its application to elastoplastic fibrous
composites. \emph{Comp. Mech.}, \textbf{7}, 13--19.






\end{thebibliography}
\end{document}